\documentclass{raa}

\usepackage{array}
\usepackage{booktabs}
\usepackage{graphicx, times}
\usepackage{natbib}
\usepackage{amssymb,amsmath}
\usepackage{bm}
\usepackage{xcolor}

\begin{document}
\title{Stellar Dynamical Modeling - Accuracy of 3D Density Estimation for Edge-on Axisymmetric Galaxies}


\author{Richard J. Long\inst{1,2}, Shengdong Lu\inst{1}, Dandan Xu\inst{1}}


\institute{Department of Astronomy, Tsinghua University, Beijing 100084, China; {\it rjlastro@yahoo.com; Orcid:0000-0002-8559-0067} \\
        \and
           Jodrell Bank Centre for Astrophysics, Department of Physics and Astronomy, The University of Manchester, Oxford Road, Manchester M13 9PL, UK\\
           }

\date{Received~~2020 month day; accepted~~2020~~month day}

\newcolumntype{M}[1]{>{\centering\arraybackslash}m{#1}}

\abstract{
From Rybicki's analysis using the Fourier slice theorem, mathematically it is possible to reproduce uniquely an edge-on axisymmetric galaxy's 3D light distribution from its 2D surface brightness. 
Utilizing galaxies from a cosmological simulation, we examine the ability of Syer and Tremaine's made-to-measure method and Schwarzschild's method for stellar dynamical modeling to do so for edge-on oblate axisymmetric galaxies. Overall, we find that the methods do not accurately recover the 3D distributions, with the made-to-measure method producing more accurate estimates than Schwarzschild's method. Our results have implications broader than just luminosity density, and affect other luminosity-weighted distributions within galaxies, for example, age and metallicity.
}

\authorrunning{Richard J. Long et al.}            
\titlerunning{Accuracy of 3D Density Estimation}  

\maketitle


\section{Introduction}
\label{sec:intro}

At optical wave lengths, our understanding of external galaxies is limited by our observational capabilities: we are unable to observe individual stars in these galaxies and so are reliant on projected brightness data and decoding blended light from many stars.  
GAIA-like observations of our galaxy \citep{GAIA2016} with stellar data including 3D positions and velocities are a luxury we do not have with external galaxies.
In addition to a galaxy's surface brightness, our main observations comprise 2D on-sky positions, with integrated line-of-sight (los) photometric spectral data leading to los kinematics and chemistry. These observations are taken with optical telescopes mounted with integral field units (IFUs) implementing integral field spectroscopy.  
Galaxy surveys delivering IFU data for individual external galaxies include ATLAS$^{3D}$ \citep{AtlasI}, CALIFA \citep{Sanchez2012}, MaNGA \citep{Bundy2015}, SAMI \citep{SAMI2015}, and MUSE \citep{Bacon2017}.

From observations, we are unable to recover the full phase space data to produce accurate, full dimensional models of observed galaxies.  Except in a few special cases, data deprojections are in general not unique, and, mathematically, this is impossible to overcome: in linear algebra terms, projection matrices are singular.
Regardless of whether models are able to reproduce the observations or not, model solutions can not be unique, and so our understanding of the observed galaxies is inherently degenerate as a consequence.  This point regarding deprojection is well-made in the introduction section of \cite{Cappellari2020}, and in \cite{Vasiliev2020}.
The introduction of statistical data from, say, cosmological simulations or stellar population synthesis may assist us by helping to eliminate non-physical models but the resultant solutions remain unable to overcome the deprojection non-uniqueness.  As such the models indicate what galaxies may be like but can not be considered as accurate.
We mentioned that there are special cases when deprojection is mathematically possible. The first case is for a spherical system using Eddington's formula \citep{BT2008}.  The second is for oblate axisymmetric systems and is the subject of this paper.

\cite{Rybicki1987} makes it clear that an axisymmetric galaxy's 2D surface brightness can only be deprojected to give the 3D light distribution if the galaxy's symmetry axis is in the plane of the sky.  In other words, deprojection is only possible if the galaxy is viewed edge-on, that is when the inclination angle is $90 ^{\circ}$ degrees. \cite{Gerhard1996} expand on this work, introducing 'konus' densities and demonstrating that an infinite number of plausible 3D densities result in the same surface brightness when the inclination is not edge-on.  A key observation  in \cite{Gerhard1996} is that triaxial galaxies must be similarly affected.  In this paper we will not investigate konus densities further, as was performed by \cite{VDBosch1997} or \cite{Zhao2000}, for example.  It should be noted that, even though \cite{Rybicki1987} is framed in terms of light, its findings apply to any axisymmetric feature of a galaxy which is observed, or constructed, as a 2D projection of a 3D distribution, for example, age or metallicity perhaps.  Similarly, the findings of \cite{Gerhard1996} can be extended to any triaxially distributed features.  \cite{deNicola2020} in a recent paper investigated a non-parametric approach to triaxial deprojection: we do not take a similar route.

In this paper, we use two stellar dynamical modeling methods, the orbit method of \cite{Schwarz1979} and the made-to-measure (M2M) particle method of \cite{Syer1996}, with our implementations being based on \cite{Long2018} for Schwarzschild's method, and \cite{Long2016} for M2M.  Our research objectives are to investigate whether or not these methods and their implementations do recover 3D axisymmetric spatial distributions from 2D projections for edge-on axisymmetric galaxies, and to investigate the extent to which recovery degrades with inclination.  In so doing, we will use matter distributions taken from a cosmological simulation.

The structure of our paper is as follows.  Section \ref{sec:approach} describes at a top level the approach we will take to our investigations.  Section \ref{sec:data} describes our galaxies and data.  Any relevant theory and descriptions of the methods are in Section \ref{sec:theory}. Note that we will not describe Schwarzschild's method and the M2M method in any great detail but refer the reader elsewhere.  Our results and any ensuing discussions are in Sections \ref{sec:results} and \ref{sec:discussion}, with our conclusions in Section \ref{sec:conclusions}.

\section{Approach} 
\label{sec:approach}

Our objectives imply that we require oblate axisymmetric galaxies with known 3D density distributions to be available to the investigation.  Known distributions mean that use of real galaxies is ruled out, and so we select appropriate axisymmetric galaxies from the IllustrisTNG project's TNG100 cosmological simulation \citep{Nelson2019}.  Since no cosmologically simulated galaxies are truly axisymmetric (as required by Rybicki's theory), we enforce axisymmetry by using multi-Gaussian expansions (MGE, see below) of the galaxies' surface brightness maps, and model galaxies both with and without kinematic constraints.

We have chosen to use Schwarzschild's method and Syer \& Tremaine's made-to-measure method in our modeling.  Arriving at 3D density estimates from these methods is straightforward.  M2M uses a 3D particle model and so 3D density estimates could be calculated from the end of run weighted particles.  Such calculations produce a numerically noisy result which can be avoided if the 3D density estimation takes place while the particle weights are being adjusted, and is exponentially smoothed as happens for the constraints. 
For Schwarzschild's method, there are two estimation processes which can be used.  The first involves building a 3D particle from the weighted orbits as in \cite{Zhao1996} and \cite{Wu2017} and then binning particles on a 3D grid to arrive at the density estimate.  The second employs a technique not too dissimilar to that used with the M2M method.  The orbit contributions to the 3D density are collected as for other constraints but these contributions are not used when the orbit weights are calculated.  Once the weights are available, they are used with the orbit 3D density contributions to determine the density estimate.  We will demonstrate that both methods yield the same density estimate to within some tolerance.

Our assessment of deprojection accuracy is consistent with the style of operation of the modeling methods in that we use the residual maps formed by comparing model observable values with their constraining values taken from taken from our galaxy data.  The extent of their difference expressed as a percentage for all observable points is used as our accuracy measure.

Both M2M and Schwarzschild's methods can only utilize the particles or orbits determined by the initial conditions and the gravitational potential.  The question being answered by the methods is whether or not there is some weighting of the orbits that can be found to enable the model observables to match the real observations.  Changing the orbits provided implies a different answer - see \cite{Cappellari2020}.  The orbit or particle weightings are not determined by astrophysical considerations.  For example, with an implementation of Schwarzschild's method using \cite{LH1974} just which orbits are zero-weighted is determined by the non-negative least squares (NNLS) code and does not involve any knowledge as to which orbit types are astrophysically more likely.  For Schwarzschild's method, in line with \cite{Long2018}, we use the convex optimization package CVXOPT \citep{CVX} to determine the orbit weights. More information on different NLLS methods is contained  in \cite{Chen2009}.  Two methods are employed for the particle and orbit initial conditions.  We continue the use of the three integral scheme as in \cite{Long2016} and \cite{Long2018}, and also introduce an additional, observationally motivated scheme based around the Jeans equations. Both schemes are described in Section \ref{sec:ics}.

We do not use regularization \citep{Tikhonov1963} as a matter of course in our models.  We do however run a small number of models with a quadratic regularization term penalizing large, orbit or particle weights to evaluate its effect on 3D density estimates.  Using a quadratic form, as in \cite{Valluri2004}, \cite{Vasiliev2013} and \cite{Long2016}, means that the convex nature of the optimization that yields the orbit weights is maintained. The same quadratic scheme is available for use with both M2M and Schwarzschild's methods.

We endeavor to employ as much commonality as possible in our use of the Schwarzschild and M2M methods.  For a given galaxy, the gravitational potentials are the same; the data are the same apart from any mass or luminosity weighting for Schwarzschild purposes;  model bin sizes, regularization schemes, and initial conditions are also the same.

A galaxy's potential is modeled as two multi-Gaussian expansions (MGEs, see \citealt{Emsellem1994}), an axisymmetric one for the stellar matter and a spherical one for the dark matter.  As indicated earlier, we are using the stellar MGE to enforce axisymmetry on the models.  For a single galaxy, the models we run utilize these MGEs with luminosity constraints coming from the axisymmetric stellar component, and with and without kinematic constraints. The full set of models for a galaxy is
\begin{enumerate}
\item \label{itm:one} a full mass, edge-on galaxy model using both MGEs and with stellar luminosity constraints only,
\item \label{itm:two} as item \ref{itm:one} with kinematic constraints included as well as luminosity constraints,
\item \label{itm:three} as item \ref{itm:one} with the galaxy inclined to the line of sight (seven different inclinations).
\end{enumerate}
Every model comprises two parts: the first uses the known 3D stellar MGE density as a constraint to ensure that the 3D density is able to be reproduced by the modeling method; and, the second, the 2D stellar surface brightness so that the extent to which a model's 3D density estimate matches the true 3D density can be assessed.  Every model is run using the two different sets of orbit / particle initial conditions. Particularly in figures, we will refer to models with no kinematic constraints as \textit{Nokin} models, and similarly \textit{Kin} for models with kinematic constraints.

Even though we run models with and without kinematic constraints, the models without kinematic constraints are produced as an aid to understanding. It is the models with kinematics constraints that will eventually guide whether our objectives have been met or not.

\section{Galaxy Data}
\label{sec:data}

In this section, we describe how we select suitable galaxies for our investigation from the publicly released IllustrisTNG project's TNG100 simulation data \citep{Nelson2019}, and how, from those galaxies, we construct IFU-like data for modeling purposes.

Our galaxy selection strategy is designed to ensure we initially bias our selection towards oblate elliptical galaxies with sufficient mass resolution such that galaxies with possible spiral features that are largely present in star forming disk galaxies are not included in our sample.  To that end, we first select galaxies with stellar mass larger than $10^{10.5}M_\odot$, with a bulge to total luminosity ratio larger than 0.5, and with a minor to major axial ratio in the range $0.42 < c/a < 0.9$.  We then refine our selection to oblate central (not satellite) galaxies with simple kinematics, perhaps with rotation around the symmetry axis, and close alignment of the mass and kinematic axes.  Close in this context means to within 3 degrees. Our criterion for axisymmetry is that the axial ratio of the axes perpendicular to the symmetry axis (the minor axis) should be such that $b/a > 0.95$.    

Approximately 130 galaxies met our criteria, and in Table \ref{tab:closegal} we show the five galaxies we have selected for modeling. The S{\'e}rsic indices \citep{Sersic1963} of the galaxies are close to the accepted elliptical galaxy value of $4$ for all the galaxies.  Both the S{\'e}rsic indices and the hot orbit fractions are taken from \cite{Xu2019}.  Galaxy A1190 has a position angle difference $>3$ degrees and has been included deliberately.

We take $R_{hsm}$, the half stellar mass radius of a galaxy, as the galaxy's `effective radius' $R_e$ and will use it to scale distances within our models (see Section \ref{sec:common}).
For data preparation purposes, we assume that all galaxies are at redshift $z=0.01$, which means 1kpc equals 1.612 arcsec.

\begin{table}
	\centering
	\caption{TNG100 Galaxies - Selection and Modeling Data}
	\label{tab:closegal}
		\begin{tabular}{crrrrrrrrcc}
		\hline
		Galaxy Id &	Sub Id & 	$R_{hsm}$ &  $\log(M^*)$ &	 S{\'e}rsic &	 Frac. & $\Delta pa$ & 	$b/a$ & 	$c/a$ & Model size & Voronoi \\
		          &        &    (kpc)     &              &     Index  &    Hot  &            &        &       &  ($R_{hsm}$) & Cells\\
		\hline 
		A490 &318633&6.586&11.13&3.79&0.34&1.68&0.98&0.46&8  &197\\
		A1090&377539&6.354&11.00&3.67&0.29&1.33&0.95&0.64&8  &110\\
		A1190&414146&6.090&10.77&4.96&0.33&6.63&0.99&0.67&8  &173\\
		A1290&416112&5.590&10.95&4.84&0.31&1.27&0.99&0.55&9  &119\\
		A1390&443421&5.154&10.82&4.46&0.27&4.00&0.96&0.52&9  &220\\
		\hline
	\end{tabular}

\medskip
The columns are our galaxy identifier, the TNG100 sub-halo identifier, the half stellar mass radius $R_{hsm}$, the stellar mass $M_*$, the S{\'e}rsic index, the fraction of hot orbits (calculated inside 2$R_{hsm}$), the position angle difference in degrees between the mass and velocity kinematic axes, and the stellar axial ratios $b/a$ and $c/a$ in the usual notation.  The axial ratios are calculated using the inner 10kpc of data. Model sizes are in units of the half stellar mass radius $R_{hsm}$.  The Voronoi cells are used for kinematic data constraints only.
\end{table}

For all galaxies, we are concerned only with the stellar and dark matter particles.  The focus in our work is on modeling the stellar mass (light) distribution so black hole particles are ignored.  We do not utilize simulation calculated luminosity values but just use the stellar mass distribution taking the stellar mass-to-light ratio as $1$.  Gas plays no role in the modeling methods we are evaluating.  We use the virial radius ($r_{200}$ to be precise, see \cite{BT2008} section 9.2) to set the overall size of a galaxy and to identify the galaxy's particles.  No further filter on stellar particles is employed.  The dark matter component of a galaxy's gravitational potential is determined by fitting a generalized spherical NFW profile (\cite{NFW1996,NFW1997}, and see also \cite{An2013}) to the dark matter particles and then converting the profile into a multi-Gaussian expansion (MGE).  Similarly, the axisymmetric stellar potential is determined from an MGE of the galaxy's edge-on surface brightness.  Note that the MGE formalism automatically gives us axisymmetric expressions for the 3D densities.  The stellar MGEs for the galaxies we have chosen are shown in Figure \ref{fig:stellarmge}.

For modeling purposes, luminosity data are always calculated from the stellar MGEs.  Brightness data for inclined galaxies are found by adjusting the observed flattening terms within the MGE (Eq. \ref{eq:flat}) and then using the adjusted MGE to calculate the data values. 
\begin{equation}
\label{eq:flat}
q_i = \sqrt{q^2_{90} \sin^2i + \cos^2i},
\end{equation}
where $q_{90}$ is the edge-on flattening, $i$ is the galaxy's inclination to the line-of sight, and $q_i$ is the `on-sky' projected flattening.
  
Generation of IFU-like stellar kinematic data is achieved by viewing the galaxy along some desired line-of-sight (inclination), binning stellar particle data into 0.5kpc square bins, and then re-binning into Voronoi bins \citep{CapCopin2003} to achieve a minimum signal to noise, $S/N \sim 20$.  In practice, only edge-on kinematic data have been required in this research. Edge-on line-of-sight velocity maps for our selected galaxies are shown in appendix \ref{app:gdata}.  The fraction of hot orbits in Table \ref{tab:closegal} is used in assessing the suitability of the initial conditions (see section \ref{sec:ics}).  Once we have constructed the MGEs and kinematic data, we make no further use of the cosmological simulation's particles.

For the galaxies we have selected, we do not attempt to model any quantities other than luminosity and kinematics.  We note however that age or metallicity data, say, could be modeled but if they are not axisymmetrically distributed (which is likely) then the observations made in \cite{Gerhard1996} will apply and it will not be possible to estimate their 3D distributions with any accuracy.

In this investigation, we are trying to establish whether or not two general purpose modeling schemes are able to deliver models that conform to the theoretical statements in \cite{Rybicki1987}.
To assist us in developing our understanding of the deprojection capabilities of the Schwarzschild and M2M modeling schemes, we run galaxy models with and without kinematic data constraints.

\section{Theory and Methods} 
\label{sec:theory}

\subsection{General Theoretical Considerations} 
\label{sec:considerations}
The theoretical considerations behind this paper come from the following four points.
\begin{enumerate}
\item Except in a few special cases, deprojection of lower dimensional galaxy data to higher dimensions is not unique.  This means that quantities calculated from the phase space coordinates arising from the orbits and particles in the modeling methods can not be taken as accurately representing a galaxy. 
\item \label{item:theory2} One of these special cases is edge-on axisymmetric galaxies where the work of \cite{Rybicki1987}, \cite{Gerhard1996} and \cite{Kochanek1996} applies, and unique deprojections are possible mathematically.  The question being asked in this paper is whether or not Schwarzschild's method and the M2M method (and their implementations) are capable of such unique deprojections.
\item An examination of \cite{Rybicki1987} indicates that its arguments are not specific to light or mass distributions and should apply to other 3D axisymmetric distributions.    In addition the arguments in \cite{Gerhard1996} regarding 3D non-axisymmetric distributions also apply. Recovery of metallicity and age distributions, for example, would therefore also be subject to the question asked in point \ref{item:theory2}.
\item From the theory of orbits in an axisymmetric potential (section 3.2 in \citealt{BT2008}), the symmetry axis (the z-axis) is protected as if by a `centrifugal barrier'.  In the context of this investigation, it is important that this barrier is overcome by ensuring that orbits with low or zero angular momentum about the z-axis are included.  Failure to do so would mean that the modeling methods would be prevented from modeling the 3D density close to the axis.
\end{enumerate}

\subsection{Schwarzschild's Method}
\label{sec:schw}
The modeling method described in \cite{Schwarz1979} to create equilibrium triaxial stellar models has evolved over time into a method for modeling external galaxies. Our implementation in  \cite{Long2018} is influenced by the use of Gauss-Hermite coefficients as constraints in \cite{Rix1997} and by the implementation used in \cite{Cappellari2006} and \cite{Remco2008}.  The method takes a galaxy's gravitational potential and constructs a library of orbits recording the orbits' contributions to some set of constraining, luminosity and kinematic observables  These orbit contributions are then weighted in an attempt to reproduce the constraints' observational data.  We refer the reader to \cite{Long2018} for more information.

In order to use Schwarzschild's method to estimate a 3D density in the absence of a 3D density constraint, we have two options.  The first is construct a weighted 3D particle model from the orbits in Schwarzschild modeling, and then calculate a 3D density estimate from that particle model (see \citealt{Zhao1996} and \citealt{Wu2017}).  The second is motivated by the M2M estimation approach in \ref{sec:m2m} and involves collecting the orbit contributions to the 3D density but not using these contributions in the weight calculations.  Once weights have been calculated, the orbit contributions can be used with the weights to produce the density estimate.  In Section \ref{sec:schwest}, we demonstrate that the options produce very similar results.

In our models, we fix the number of orbits at 8000, and collect orbit contributions over 50 hmdtu (half mass dynamical time units) (see \citealt{BT2008}).

\subsubsection{Schwarzschild 3D Density Estimators}
\label{sec:schwest}

In previous section, we indicated that there are two means of estimating a 3D density using Schwarzschild's method.  The first requires a particle model to be created, and the second needs a slight adjustment to Schwarzschild's method. We now demonstrate the two estimation methods for the \textit{Nokin} and \textit{Kin} sets of models, using both the 3I and MDJV schemes for initial conditions described in Section \ref{sec:ics}.

For the particle model, we must first run Schwarzschild's method to produce the orbit weights.  The particle model is then created by taking each orbit and dividing the orbit into a number $N$ of equal time intervals.  After each time interval we create a particle of weight $w_j / N$ at the position reached along the orbit ($w_j$ is the weight of orbit $j$).  Binning these particles allows the 3D density to be estimated. We take $N=1000$ giving $8 \times 10^6$ particles per model.

\begin{figure}[h]
    \centering
    \caption{Comparison of Schwarzschild 3D Density Estimators}
	\label{fig:schwest}
    \begin{tabular}{cM{65mm}M{65mm}M{65mm}}
    		& \textit{3I Initial Conditions} & \textit{MDJV Initial Conditions} \\
      	\textit{Nokin} & \includegraphics[width=65mm]{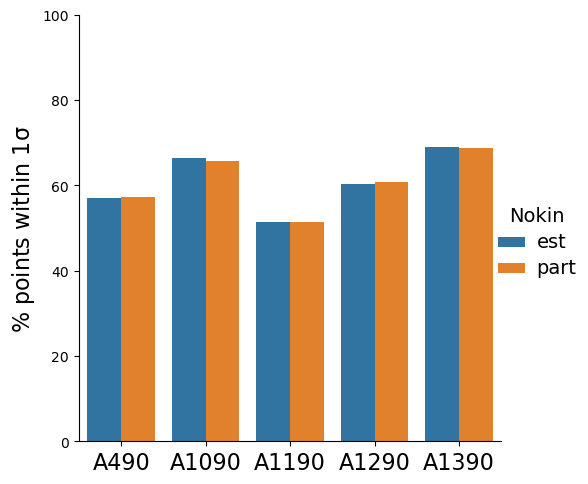}  &  \includegraphics[width=65mm]{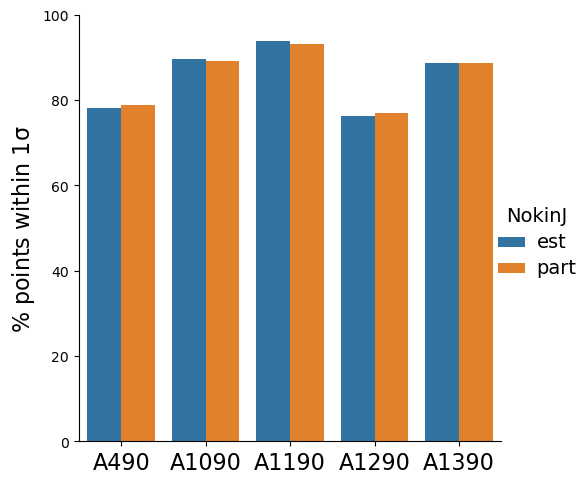}  \\
      	\textit{Kin}   & \includegraphics[width=65mm]{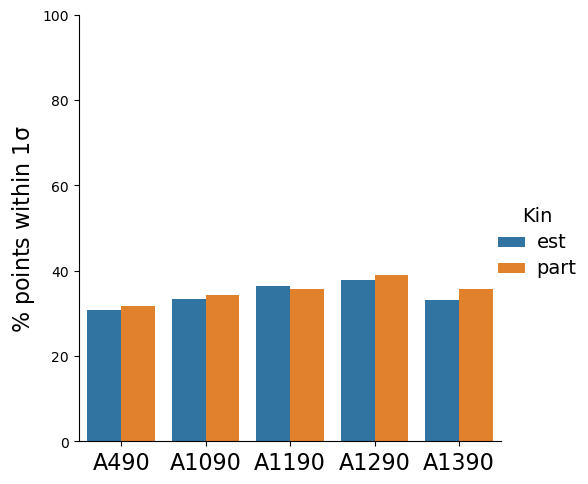}    &  \includegraphics[width=65mm]{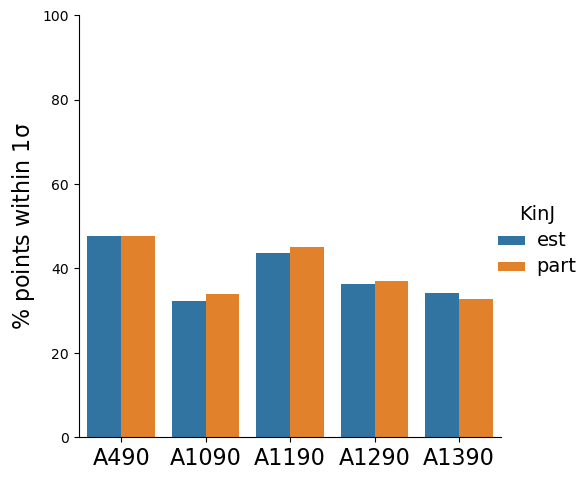}  \\
    \end{tabular}
    
\medskip  
Comparison between the two methods for Schwarzschild 3D density estimation (Section \ref{sec:schwest}).  Blue indicates the embedded Schwarzschild estimator values and orange the particle method values.  The individual plots show the $1\sigma$ residual percentages - see Eq. (\ref{eqn:ptresid}). Overall, there are no significant differences between the two estimation methods.
\end{figure}

In Figure \ref{fig:schwest} we compare the two estimation methods.  Results using 3I initial conditions are in the first column, and MDJV in the second column, with the individual plots showing the $1\sigma$ residual percentages - see Eq. (\ref{eqn:ptresid}).  \textit{Nokin} models for the galaxies are in the first row, with \textit{Kin} models in the second row.  As can be seen, the two methods yield very similar results with typically the difference being $\approx 1\%$.  For operational convenience, we choose to use only the Schwarzschild based estimator in analyzing our results.

\subsection{Made-to-measure Method}
\label{sec:m2m}
The M2M method described in \cite{Syer1996} takes a system of weighted particles and orbits them in a gravitational potential representative of a galaxy.  While the particles are being orbited their weights are adjusted so that various observationally based constraints are met.  The following papers have all contributed in some way to the development or use of the of the method: \cite{DL2007}, \cite{Dehnen2009}, \cite{Long2010, Long2012}, \cite{Morganti2012}, \cite{Hunt2013}, \cite{Malvido2015}, \citet{Portail2015}, \cite{Long2016} and \cite{Bovy2018}. There are slight variations in the low level designs and implementations of the method.  We choose to follow \cite{Long2016}, and refer the reader to that paper for more information about the method and our implementation. 

In order to use M2M to estimate a 3D density in the absence of a 3D density constraint, we include in our M2M implementation `estimator' objects.  These objects act similarly to constraints in the sense that model estimates are calculated as the weighted particles are orbited and are put through the exponential smoothing process, but they are not included in the particle weight adaption mechanism.  A similar technique was used in \cite{Morganti2013} to estimate the $\beta (r)$ velocity dispersion anisotropy parameter.

M2M has substantially more hyper-parameters available to tune the method than Schwarzschild's method.  We fix the number of particles we use to $2 \times 10^5$ but allow some variation in model durations and the value of the $\epsilon$ parameter (which controls the rate of weight adaption) so that $\chi^2$ per bin $\approx 1$ for all observational constraints. Typically, durations are $200$ time units for models with MDJV initial conditions, and $300$ units for 3I models.  The $\epsilon$ parameter is usually either $5 \times 10^{-5}$ or $1 \times 10^{-4}$.  With M2M, we must specify the initial particle weights (there is no requirement to do so in Schwarzschild's method).  In line with earlier modeling \citep{Long2010, Long2016}, we choose to set their fractional values to $1/N$ where $N$ is the number of particles used.
Given that only the \textit{Kin} models have more than one observational constraint, apart from for these models, automatic numerical balancing in the weight adaption equation is turned off.  Overall, these parameters also enable us to achieve weight convergence of $\approx 95\%$ for 3I models and $\approx 98\%$ for MDJV models (see Section \ref{sec:ics} for an explanation of 3I and MDJV).

\subsection{Commonality} 
\label{sec:common}

In order to facilitate comparison between our chosen dynamical modeling methods, we will seek to use as much commonality as possible in our usage of the methods.  In this section, we identify  the common areas.  The units we use are the same as in \cite{Long2016} and \cite{Long2018} and are effective radii for length, $10^7$ years for time, and mass in units of the solar mass $M_\odot$ with luminosity similarly so.

The galaxy gravitational potentials are constructed from MGEs of the stellar and dark matter masses with a stellar mass-to-light ratio of $1$.  Even though the IllustrisTNG galaxies include black hole particles, we do not utilize them in our modeling: most models do not involve kinematic constraints.  Having determined the potential, the initial conditions (spatial and velocity) are constructed for our particles and orbits.  The only difference here is that the number of orbits used in Schwarzschild's models (8000) is considerably smaller than the number of particles used in the M2M models ($2 \times 10^5$).  The methods for setting the initial conditions are described in the next section, Section \ref{sec:ics}.

The luminosity constraints are 2D surface brightness and 3D luminosity density calculated from the stellar MGEs.  Error terms are numerically set as $10$ percent relative errors.  Surface brightness data are held on a polar $(R,\phi)$ $(16,16)$ grid giving $256$ cells, and luminosity density data on an axisymmetric $(r,\theta)$ $(12,25)$  half grid with $300$ cells.  Both the polar and axisymmetric grids are logarithmic in radius.
Kinematic constraints are used in Gauss-Hermite coefficient $h_1$ to $h_4$ form (\citealt{Marel1993}; \citealt{Gerhard1993}), and are symmetrized as appropriate to axisymmetry before modeling  (for example, \citealt{Cappellari2006}, \citealt{Remco2008}).  The associated Voronoi cells (see Table \ref{tab:closegal}) are used directly in the modeling with no attempt made to interpolate kinematic data onto a regular grid, say.    In addition, we use a sum of weights constraint (equal to 1) for all our models.
The modeling methods differ in the way kinematic constraints are dealt with: Schwarzschild's method requires surface luminosity times the constraint values whereas M2M does not.

For 3D luminosity density estimation (as distinct from 3D density as a constraint), we also employ an axisymmetric $(r,\theta)$ $(12,25)$  half grid with $300$ cells.

\subsection{Initial Conditions}
\label{sec:ics}

We employ two different schemes for creating initial conditions for the orbits / particles.  The first is a well established method employing sampling from the three dimensional space of the isolating integrals of motion for an axisymmetric system (\citealt{Cappellari2006}; \citealt{Remco2008}; \citealt{Long2016}), and the second is concerned with sampling observational space and using the Jeans equations (similar to that used in \citealt{Long2010, Long2012}).  

The three integral scheme uses sampling of $(E, L_z, I_3)$, energy, angular momentum about the symmetry axis z, and a surrogate third integral, to set the initial positions and velocities.  We implement the scheme in gridless form as in \cite{Long2012}.  For convenience we refer to this scheme as the 3I scheme.

The second scheme is in two parts: first, the initial spatial positions are set in such a way that the distribution profile matches that of the surface brightness MGE; and second, the initial velocities are determined by the solving the axisymmetric isotropic Jeans equations at each initial spatial position and using Gaussian sampling to set the actual velocities.  Given the gravitational potential and the initial conditions, it is straightforward to calculate the energy $E$ and angular momentum $L_z$ for use in later model analyses.   We refer to this scheme as the MDJV (match density, Jeans velocities) scheme.

\begin{figure*}
    \centering
    \caption{Comparison of 3I and MDJV Initial Conditions Schemes}
	\label{fig:icsELz}
    \begin{tabular}{cc}
        \includegraphics[width=70mm]{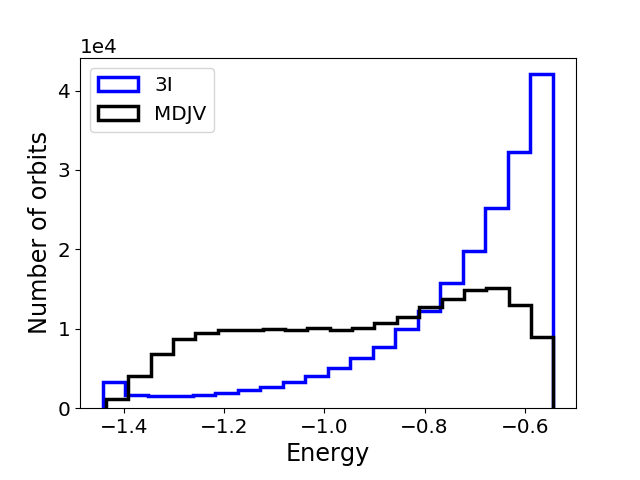} & \includegraphics[width=70mm]{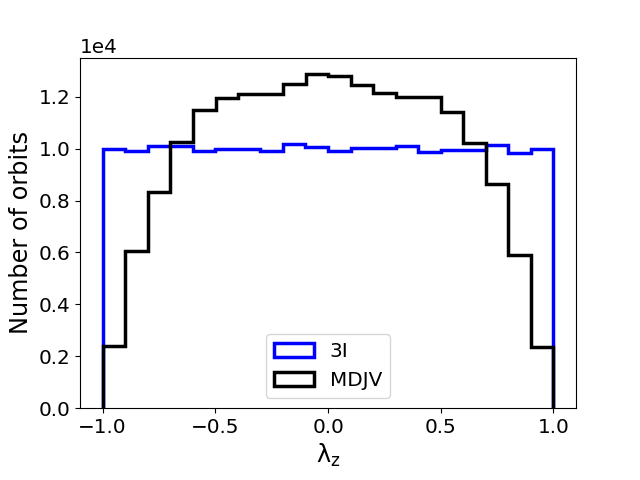} \\
    \end{tabular}

\medskip
A comparison of the 3I and MDJV initial conditions schemes (Section \ref{sec:ics}) showing how the orbit (particle) energy and angular momentum distributions differ.  We do not plot angular momentum directly but use the circularity measure $\lambda_z$ instead.  From the plots, it can be seen that the MDJV (match density, Jeans velocities) scheme generates more orbits with $\lambda_z \approx 0$ than the 3I (three integral) scheme, and this is important when analyzing our results in Section \ref{sec:results}. The precise $\lambda_z \approx 0$ difference varies for each galaxy.  The data for the plots come from the M2M \textit{Nokin} models for galaxy A1390.
\end{figure*}

As may be seen from Figure \ref{fig:icsELz}, the two schemes result in different energy and angular momentum profiles for the orbits / particles, and these different profiles will influence the results from our modeling.
We do not mix the two schemes, or add in additional orbits as other researchers have sometimes done, for example \cite{Remco2008}.  We do ensure that the two schemes share a common lower angular momentum bound ($10^{-7}$ in internal units).

Table \ref{tab:hotorbits} shows the percentages of hot orbits generated by the initial conditions schemes.  Hot orbits are those with $|\lambda_z| < 0.25$ where $\lambda_z$ is the orbit circularity measure defined as the ratio of the orbit's angular momentum to the maximum angular momentum allowed by the orbit's energy.  The 3I values of $25\%$ are to be expected given the way the 3I scheme function (see Figure \ref{fig:icsELz}, right hand plot).  The 3I scheme in general underestimates the hot orbit values from the simulation particle data by $\approx 6\%$, with the MDJV scheme matching the data values to within $\approx \pm 3\%$.
\begin{table}[h]
	\centering
	\caption{Hot Orbits Comparison}
	\label{tab:hotorbits}
		\begin{tabular}{cccc}
		\hline
		Galaxy Id &	   3I  & 	MDJV  &  Data  \\
		\hline 
		A490   & 25\% & 28\% & 34\%\\
		A1090  & 25\% & 32\% & 29\%\\
		A1190  & 25\% & 36\% & 33\%\\
		A1290  & 25\% & 28\% & 31\%\\
		A1390  & 25\% & 31\% & 27\%\\
		\hline
	\end{tabular}

\medskip
A comparison of the 3I and MDJV initial conditions schemes (Section \ref{sec:ics}) showing the percentage of hot orbits generated by each scheme.  The columns from left to right are the galaxy identifier, and then the percentages of hot orbits from the 3I and MDJV initial conditions and from the TNG100 stellar particle data.  Hot orbits are defined as those with $|\lambda_z| < 0.25$ where $\lambda_z$ is the circularity measure. The particle data values are taken directly from Table \ref{tab:closegal}.
\end{table}

M2M is more flexible than Schwarzschild's method in that the particle weights are available from the start of a modeling run.  As indicated in Section \ref{sec:m2m}, we use the same initial value for all particles.  It is quite possible to use alternative schemes, and this we have experimented with by setting initial weights using the surface luminosity at a particle's on-sky position. Having run models using both sets of initial conditions (3I and MDJV) with and without kinematic constraints, we find that our alternative scheme has little impact ($\pm2\%$) on 3D density estimates.
The determination of initial conditions uses random numbers, and, as a consequence, initial conditions are subject to stochastic variations.  We do not attempt to assess the impact of these variations in this work.

\section{Results}
\label{sec:results}

Before looking at specific modeling runs, we first describe how in general we assess our results.  As stated earlier each run is in 2 parts.  The first part is concerned with checking that, for a given model configuration, the modeling method implementation we are using is able to reproduce the 3D density when constrained to do so.  If it is unable to do this, we must stop and investigate.  The second part is concerned with running the same model configuration again but this time constrained by the 2D density (and, in later models, perhaps kinematics data as well).  What we are looking for this time is for all the constraints to be met, and to see how well the modeling method is able to estimate the 3D density distribution when it is not constrained to do so.  If the estimate is not acceptable, then we investigate why this is the case, using comparisons with the output from the first part in the diagnostic process.

\begin{figure*}
    \centering
    \caption{Examples of 3D Density Estimator Point Residual Maps}
	\label{fig:ptresid}
    \begin{tabular}{cc}
        \includegraphics[width=70mm]{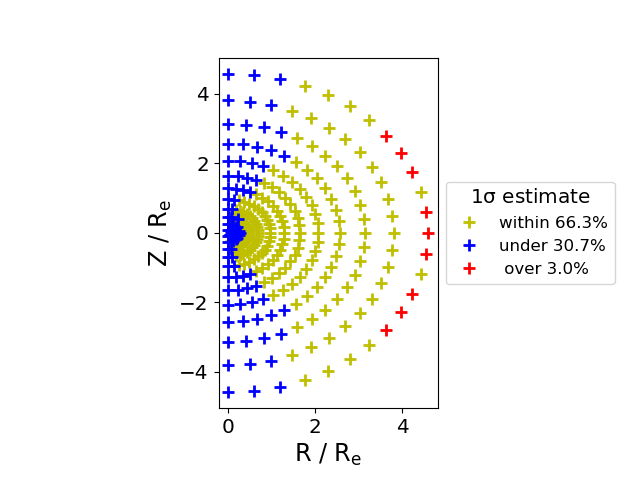} & \includegraphics[width=70mm]{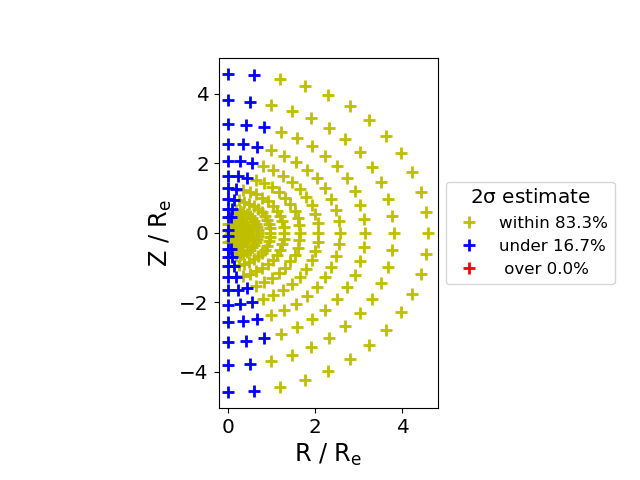} \\
    \end{tabular}

\medskip
Examples of 3D density estimator point residual maps ($1\sigma$ and $2\sigma$) taken from a Schwarzschild \textit{Nokin} model of A1090 using 3I initial conditions (Section \ref{sec:results}). Yellow points are within the error terms ($\alpha E_j$ in Eq. \ref{eqn:ptresid}), over-estimated points are red, and under-estimated points are blue.  There are $300$ points in total.  The maps show significant under-estimation along the z-axis, and minor over-estimation around the boundary.  The equivalent $1\sigma$ contour residual map is shown in Figure \ref{fig:tradres}.
\end{figure*}

\begin{figure*}
    \centering
    \caption{Example of 3D Density Estimator Contour Residual Map and Histogram}
	\label{fig:tradres}
    \begin{tabular}{cc}
        \includegraphics[width=70mm]{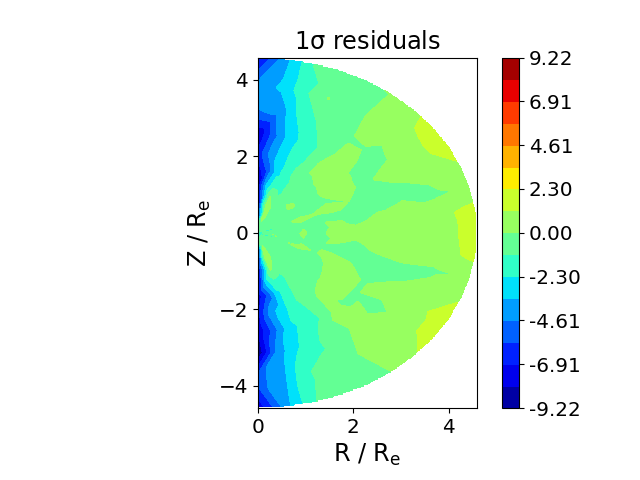} & \includegraphics[width=70mm]{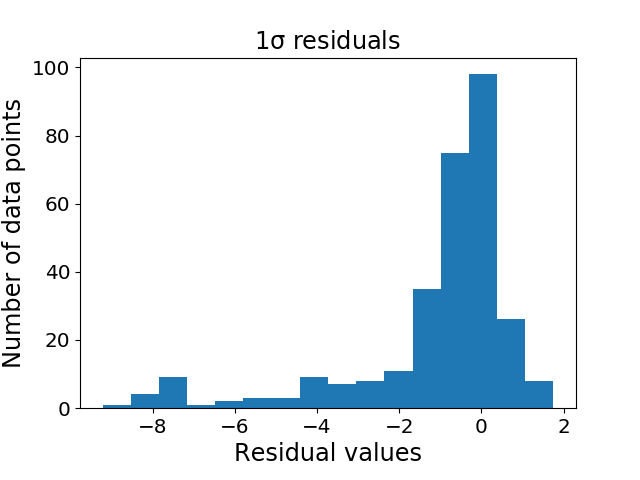} \\
    \end{tabular}

\medskip
The $1\sigma$ contour residual map equivalent to the left hand plot in Figure \ref{fig:ptresid}.  Residuals have been normalized by the error terms. The over-estimation around the boundary is less immediately visible than in Figure \ref{fig:ptresid}.
\end{figure*}

In more detail, for acceptable models, we want the $\chi ^2$ per bin value for each constraint to be $\approx 1$, where the $\chi ^2$ per bin value is calculated as the total constraint $\chi ^2$ divided by the number of constraint bins.  For the 3D density distribution estimator, given we know what the true 3D values are, again, we want the $\chi ^2 _{3D} \approx 1$ but this time, more importantly,  we also want to examine the residual maps on a point by point basis to understand what the modeling methods are and are not capable of achieving. $\chi^2$ per bin values alone are too coarse for our purposes. We normalize the residuals using the error terms, and color code our point residual maps using
\begin{equation}
\label{eqn:ptresid}
	| y_j - Y_j | < \alpha E_j
\end{equation}
where, for data point $j$, $Y_j$ and $E_j$ are the observed, measured or target value and its error, $y_j$ is the model estimated value, and $\alpha$ is one of [1, 1.5, 2, 3].  Points for which the inequality is true are colored yellow, over-estimated points are red, and under-estimated points are blue.  Loosely, we refer to estimates or percentages as $1$ or $2\sigma$ values (and so on) depending on the value of $\alpha$ being used. Figure \ref{fig:ptresid} shows some example plots taken from a Schwarzschild model of galaxy A1090 while Figure \ref{fig:tradres} shows the equivalent, more traditional contour residual map.

\begin{figure*}
    \centering
    \caption{Examples of Circularity Measure Plots}
	\label{fig:circexample}
    \begin{tabular}{cc}
        \includegraphics[width=70mm]{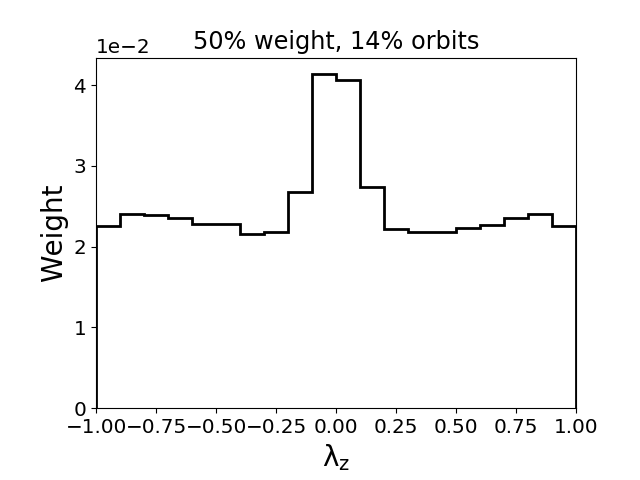} & \includegraphics[width=70mm]{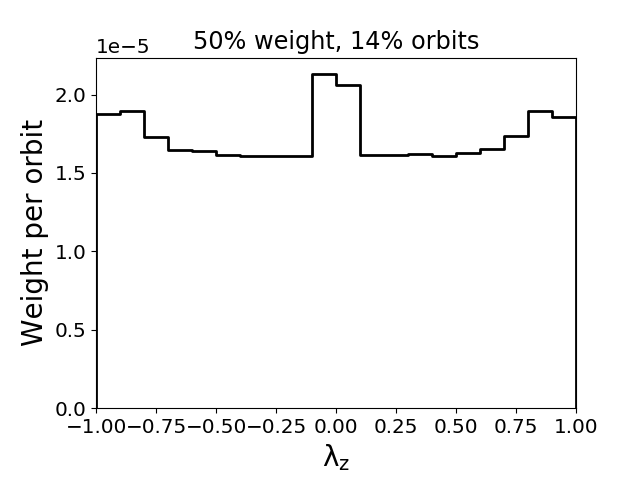} \\
    \end{tabular}
    
\medskip  
Examples of circularity measure ($\lambda _z$) plots taken from an M2M \textit{Nokin} model of A1090 using 3I initial conditions (Section \ref{sec:results}).  The left hand plot shows the weight distribution for the heaviest weighted particles making up 50\% of the total weight, and the right hand plot shows the weight per orbit.  In this case only 14\% (of $2 \times 10^5$ particles) are needed to make up 50\% of the weight (the total weight is 1 for all models).  The central peak at $\lambda_z \approx 0$ is typical of models where 3D luminosity density is being used as a constraint.
\end{figure*}

As part of our diagnostic work, we find it instructive to look at an analysis of orbit or particle weights by circularity measure $\lambda _z$.  Because Schwarzschild's method and the M2M method tend to select a small number of orbits or particles and weight them relatively highly (see, for example, \cite{Long2018}), we partition them into the heaviest orbits or particles comprising [25\%, 50\%, 75\%, 100\%] of the total weight.  Figure \ref{fig:circexample} shows some example plots taken from a \textit{Nokin} M2M model of galaxy A1090.

\subsection{3D Density Constrained Models}
\label{sec:3Dmodels}
\begin{figure}[h]
    \centering
    \caption{Edge-on Models with 3D Density Constraints}
	\label{fig:ldedgeon}
    \begin{tabular}{cM{65mm}M{65mm}M{65mm}}
         & \textit{3I Initial Conditions} & \textit{MDJV Initial Conditions} \\
        \textit{M2M}  & \includegraphics[width=65mm]{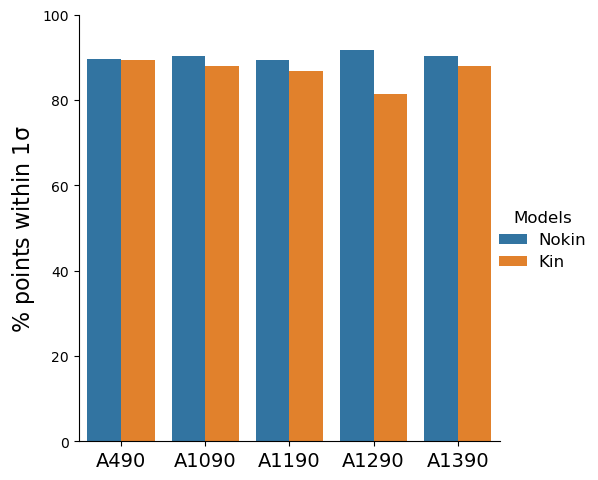}   & \includegraphics[width=65mm]{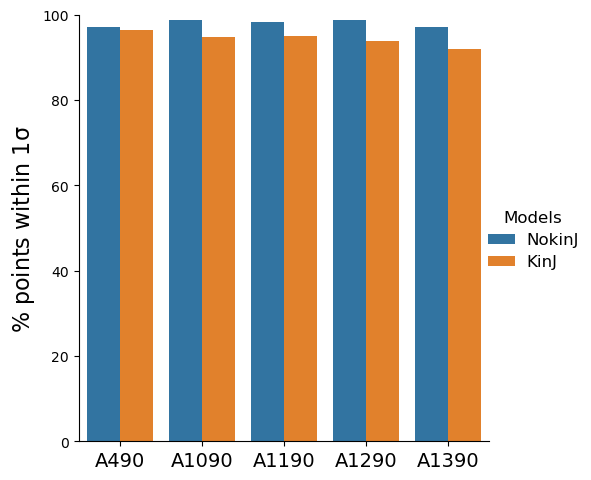} \\
        \textit{Schw} & \includegraphics[width=65mm]{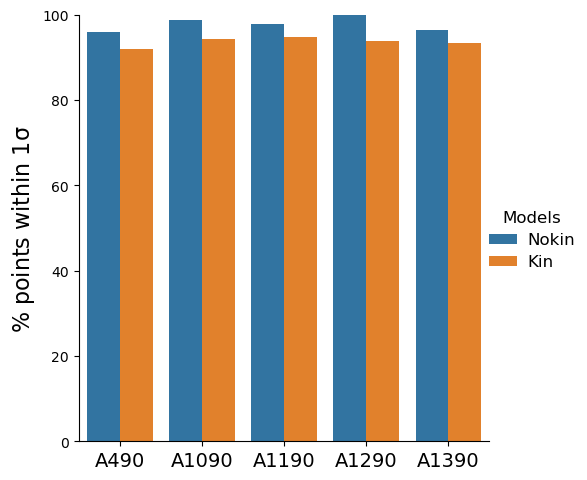}  & \includegraphics[width=65mm]{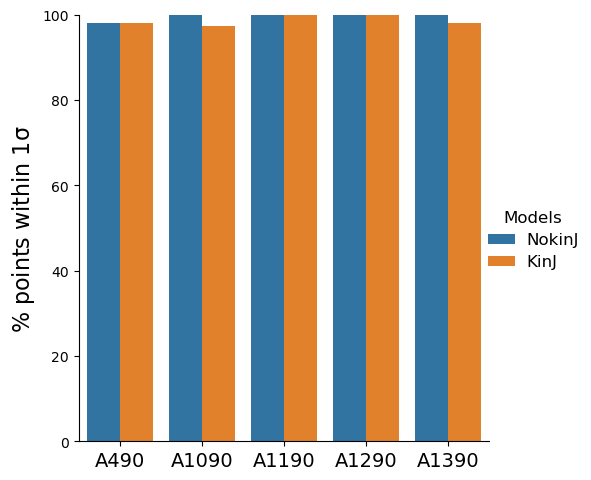} \\
    \end{tabular}

\medskip
Modeling with 3D luminosity density constraints in edge-on Schwarzschild and M2M models (Section \ref{sec:3Dmodels}). The height of the bars is the percentage of points where the model 3D density agrees with the known value to within $1\sigma$ (see Eq. \ref{eqn:ptresid} with $\alpha = 1$).  
\end{figure}

As indicated earlier, each modeling run for a galaxy is made up of two models, the first using 3D luminosity density as a constraint and the second, 2D surface brightness.  As no issues appeared with the first parts in any of the modeling groups (the 3D luminosity constraint was met), it is convenient to deal with all these models in one section. All the $\chi^2$ per bin values are $<<1$ for the Schwarzschild models and $\approx 1$ for the M2M models.  The $\chi^2$ differences come from Schwarzschild's method generally over-fitting the data if it is able to do so, and differences in the way the two methods deal with data error terms. For \textit{Kin} models, for both modeling methods and both sets of initial conditions, the Gauss-Hermite coefficients $h_1$ to $h_4$ all have     $\chi^2$ per bin values $< 1$.  We show the 3D luminosity density point residual $1\sigma$ percentages in Figure \ref{fig:ldedgeon}.  These percentages tell us how well the modeling methods are able to meet the 3D luminosity density constraints.

Summarizing Figure \ref{fig:ldedgeon}, the Schwarzschild models perform slightly better than M2M models; the MDJV initial conditions are to be preferred over 3I; and, models with kinematic constraints perform less well than models with no kinematic constraints.
It is important to understand just which orbits and particles are having a significant impact on the modeling methods' abilities to reproduce the 3D density.  Figure \ref{fig:circexample} gives a first indication with its higher-weighted $\lambda_z \approx 0$ orbits, and we will return to this point in Sections \ref{sec:2Dmodels} and \ref{sec:moredetail}. For our M2M models, any concerns regarding meeting the 3D density constraint along the z-axis (see Section \ref{sec:considerations}) have not materialized for the MDJV initial conditions.  For the 3I initial conditions, particle counts along the z-axis are low for estimation purposes but the constraint processes function as expected. For our Schwarzschild models, no z-axis concerns have been identified.

\subsection{2D Density Constrained Models}
\label{sec:2Dmodels}
\begin{figure}[h]
    \centering
    \caption{Comparison of Edge-on Models - 3D Density Estimator Percentage Residuals}
	\label{fig:sbedgeon}
    \begin{tabular}{cM{65mm}M{65mm}M{65mm}}
    		& \textit{3I Initial Conditions} & \textit{MDJV Initial Conditions} \\
        \textit{M2M}  & \includegraphics[width=65mm]{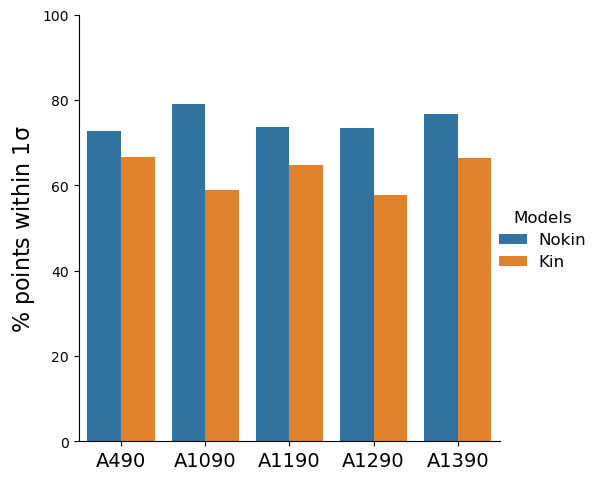}   & \includegraphics[width=65mm]{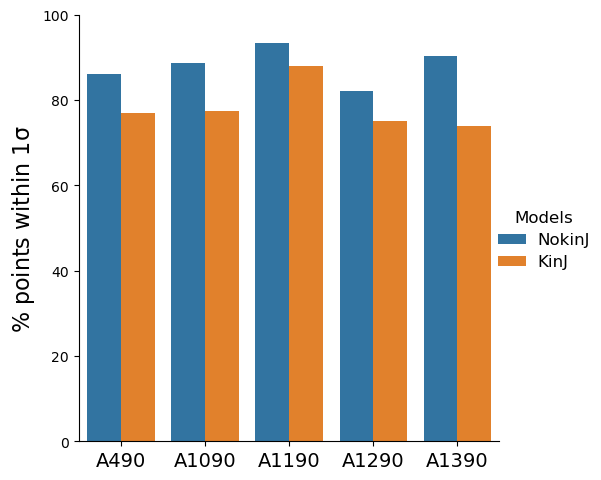} \\
        \textit{Schw} & \includegraphics[width=65mm]{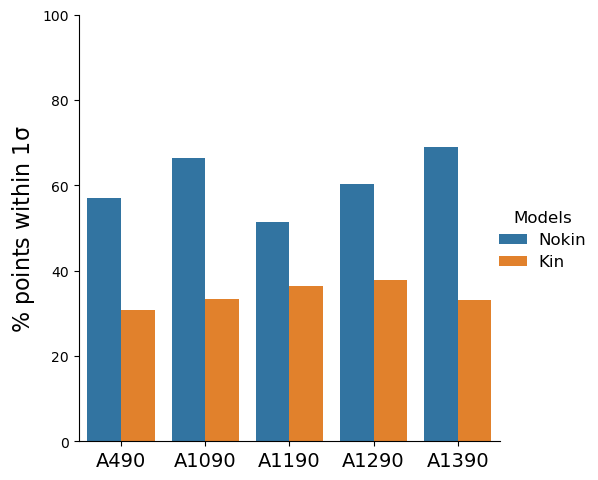}  & \includegraphics[width=65mm]{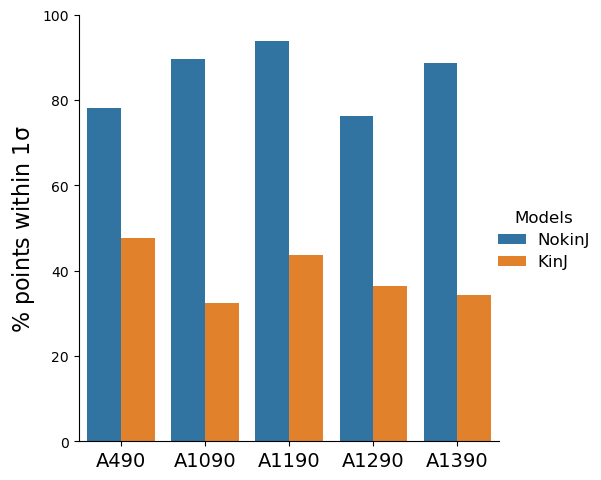} \\
    \end{tabular}

\medskip
Percentage residuals from 3D luminosity density estimation using edge-on models (Section \ref{sec:2Dmodels}). The models on the left use the 3I scheme for initial conditions while the models on the right use the MDJV scheme. The height of the bars is the percentage of points where the estimated 3D density agrees with the known value to within $1\sigma$ (see Eq. \ref{eqn:ptresid} with $\alpha = 1$).  Overall, the plots show that recovery of the 3D luminosity density improves when MDJV initial conditions are used, but decreases with the introduction of kinematic constraints.  
\end{figure}

The edge-on models in this section contribute towards points 1 and 2 from the Approach, Section \ref{sec:approach}.  Only models using surface brightness as a 2D luminosity constraint are described in this section: 3D luminosity constraints are covered in Section \ref{sec:3Dmodels}.  In all models, the surface brightness $\chi^2$ per bin values are $ < 1$.  In the \textit{Kin} models, the Gauss-Hermite coefficients $h_1$ to $h_4$ all have $\chi^2$ per bin values $< 1$.  In this respect all the models, regardless of initial conditions, are able to reproduce their constraining observables successfully.  In Figure \ref{fig:sbedgeon}, we show how well the models have been able to estimate the 3D luminosity densities of our galaxies.  The figure shows that the accuracy of the model 3D luminosity density estimates improves when the MDJV scheme for initial conditions is used, but decreases with the introduction of kinematic constraints.  Overall, the better results are achieved using M2M.

\begin{figure}[h]
    \centering
    \caption{Galaxy A1090 Point Residuals}
	\label{fig:1090pres}
    \begin{tabular}{cM{65mm}M{65mm}M{65mm}}
    		& \textit{3I Initial Conditions} & \textit{MDJV Initial Conditions} \\
    		\multicolumn{3}{c}{\textit{M2M Models}}\\
        \textit{Nokin} & \includegraphics[width=65mm]{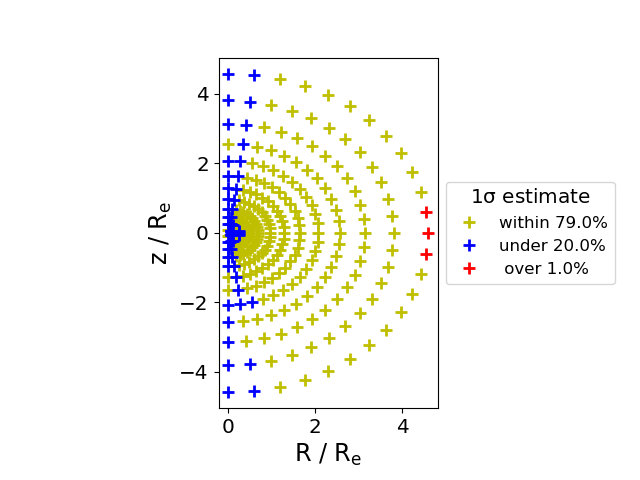}  & \includegraphics[width=65mm]{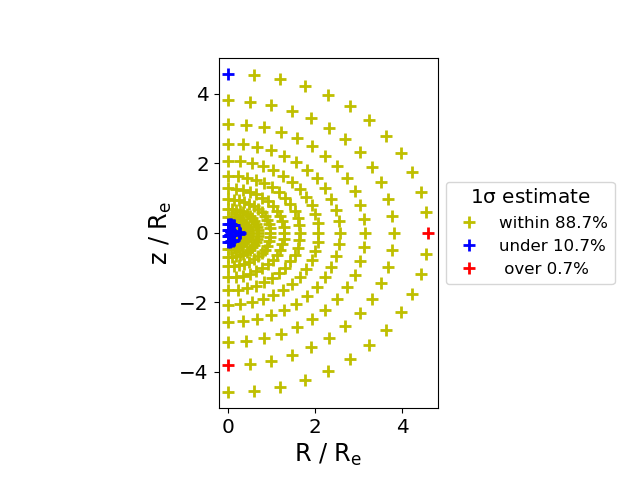} \\
        \textit{Kin} & \includegraphics[width=65mm]{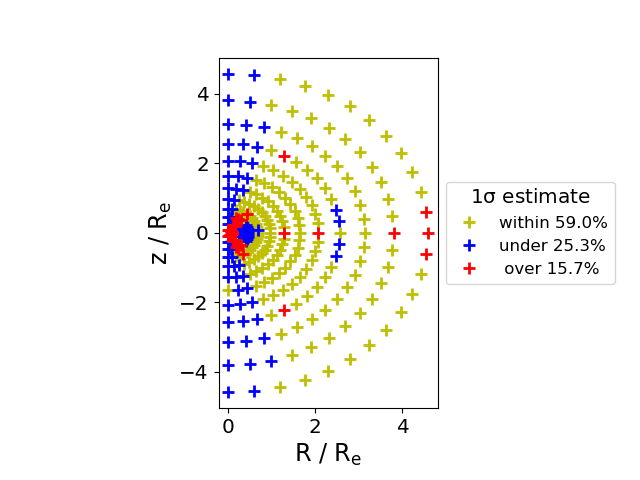}  & \includegraphics[width=65mm]{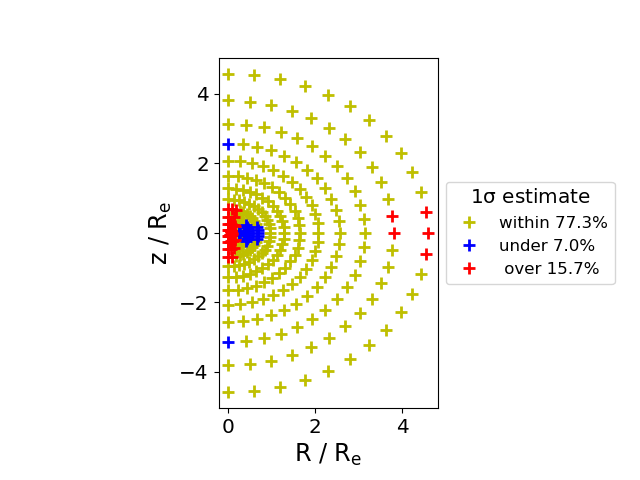} \\
        \multicolumn{3}{c}{\textit{Schwarzschild Models}}\\
        \textit{Nokin} & \includegraphics[width=65mm]{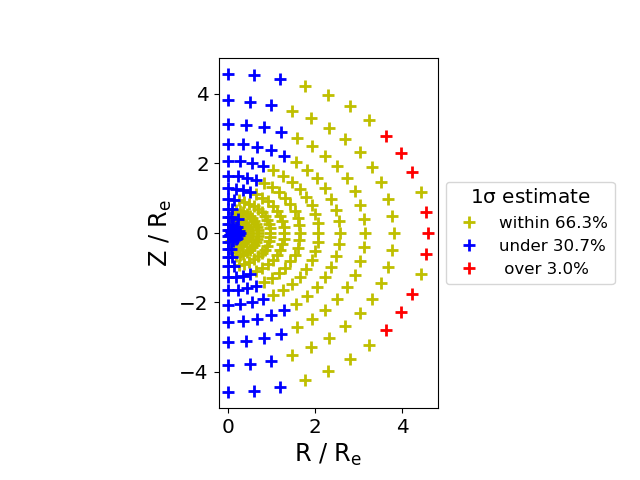}  & \includegraphics[width=65mm]{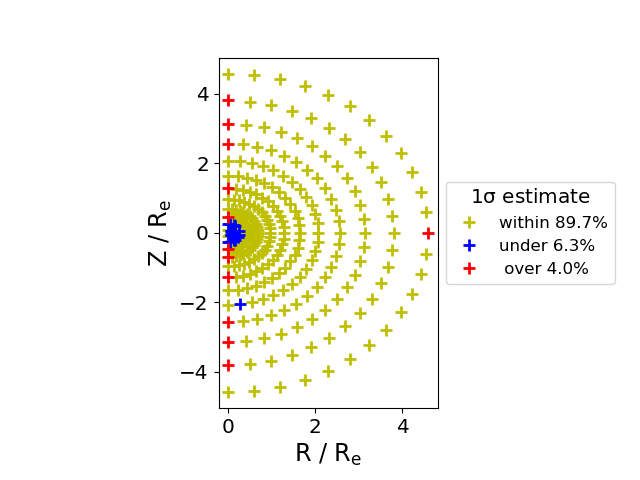} \\
        \textit{Kin} & \includegraphics[width=65mm]{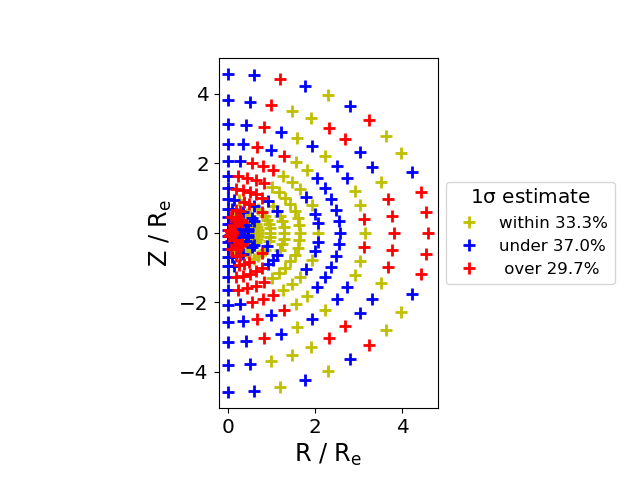}  & \includegraphics[width=65mm]{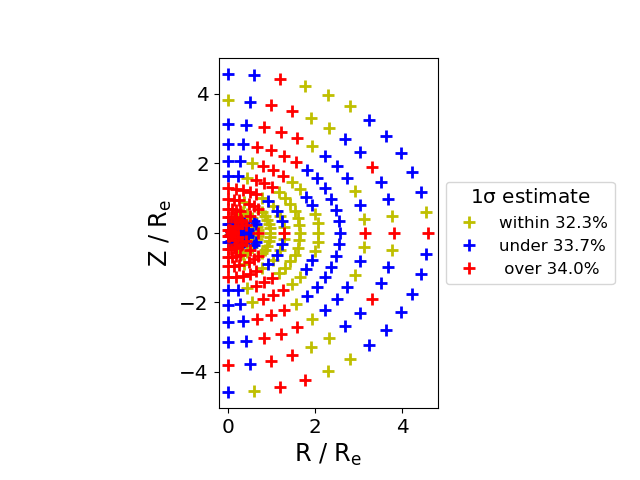} \\
    \end{tabular}

\medskip
Point residual plots for galaxy A1090 (Section \ref{sec:2Dmodels}). The \textit{Nokin} models use luminosity constraints only while the \textit{Kin} models use both luminosity and kinetic constraints.  Overall, the plots show that recovery of the 3D luminosity density along or close to the z-axis improves when MDJV initial conditions are used, but decreases for both sets of initial conditions with the introduction of kinematic constraints.  
\end{figure}

Reviewing 3D density estimates across the models at the individual point level, it is clear that estimates along or close to the z-axis are not accurately reflecting their true values well. As with the percentage residuals analysis above, the MDJV initial conditions have the better z-axis performance, and the introduction of kinematic constraints reduces the accuracy of the estimates.  In Figure \ref{fig:1090pres}, we illustrate the z-axis issue using point residual plots from galaxy A1090 M2M modeling runs.
We assess the comparative success of the M2M method as being due to a combination of factors including the number of orbits, the initial conditions, the ability to specify initial weights, and the mechanism used to determine particle weights.

\subsection{Weights Analysis}
\label{sec:moredetail}
\begin{figure}[h]
    \centering
    \caption{Weight Comparison between Luminosity Constraints for A1390 \textit{Nokin} Models}
	\label{fig:1390ldvssb}
    \begin{tabular}{cM{65mm}M{65mm}M{65mm}}    		
        \textit{M2M} & \includegraphics[width=65mm]{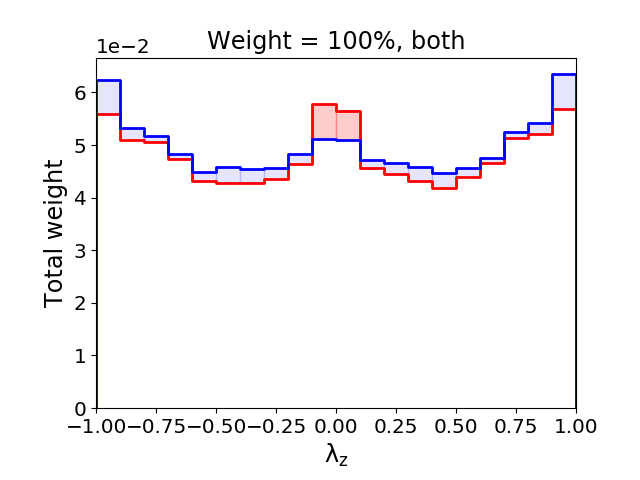}  & \includegraphics[width=65mm]{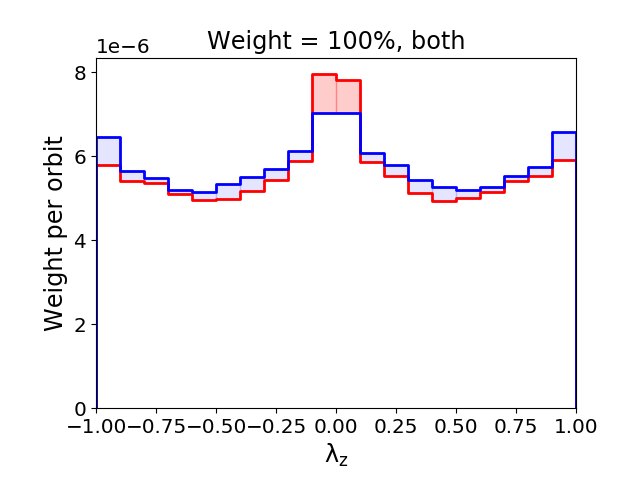} \\
        \textit{Schw} & \includegraphics[width=65mm]{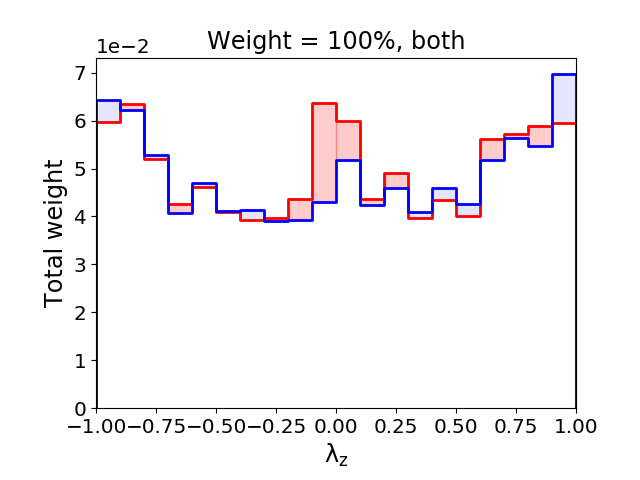}  & \includegraphics[width=65mm]{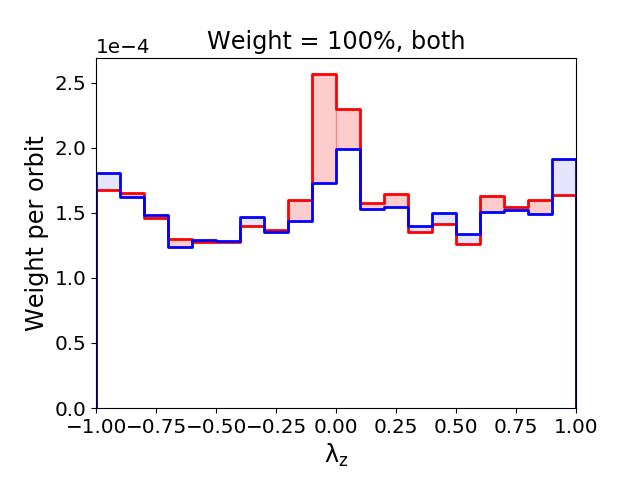} \\
    \end{tabular}

\medskip
Overall weight and weight per orbit comparisons showing differences in the weighting of low $|\lambda_z|$ orbits between the 3D and 2D luminosity constraints (Section \ref{sec:moredetail}). The red line and shading indicate the 3D luminosity density constraint profile, and the blue line and shading are the same for 2D surface brightness. The shading indicates which of the two constraints produces higher values in a given $\lambda_z$ interval  The higher $\lambda_z \approx 0$ weight levels for luminosity density (red) are a direct consequence of using a 3D constraint.
\end{figure}

\begin{figure}[h]
    \centering
    \caption{Individual Weight Comparison between Luminosity Constraints for A1390 \textit{Nokin} Models}
	\label{fig:1390ldsbweights}
    \begin{tabular}{ccc}
    		\textit{M2M Model} & \textit{Schwarzschild Model}\\
        \includegraphics[width=70mm]{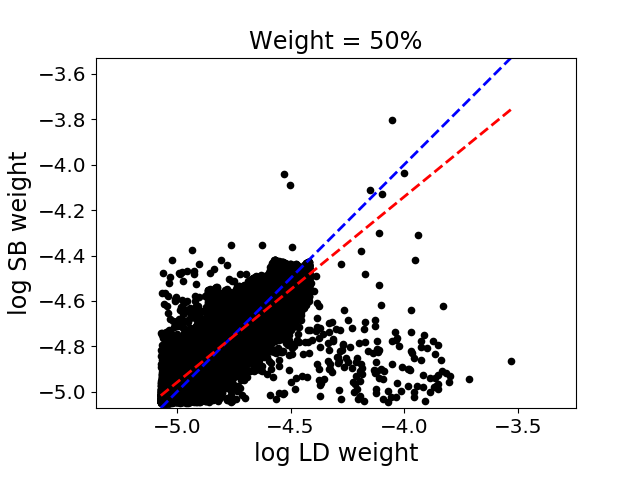} &      \includegraphics[width=70mm]{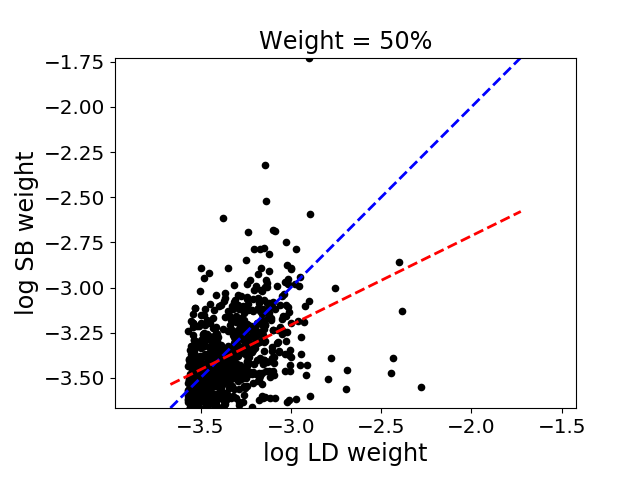} \\
    \end{tabular}

\medskip
Individual weight comparison between luminosity constraints (Section \ref{sec:moredetail}). Each plot shows the common heaviest orbits which make up 50\% of the total weight.  The blue dashed line indicates equality of the luminosity density and surface brightness weights while the red line is the best-fit to the two sets of model weights.
\end{figure}

Comparing circularity measure ($\lambda_z$) plots between the models in Section \ref{sec:2Dmodels} (2D surface brightness constraints) with the models in the Section \ref{sec:3Dmodels} (3D luminosity densities constraints) is instructive. Figure \ref{fig:1390ldvssb} contains the relevant plots, while Figure \ref{fig:1390ldsbweights} shows the difference in weights created by using the two different constraints.  We restrict the figures to just one galaxy, A1390, and show results from its \textit{Nokin} models.  The other galaxies have similar results.

The weight comparison in Figure \ref{fig:1390ldvssb} shows differences in the weighting of low $|\lambda_z|$, hot orbits between the 3D and 2D luminosity constraints.  The lower, low $|\lambda_z|$ weight levels for surface brightness are a direct consequence of only using a 2D constraint, which results in the brightness models being unable to model the 3D density along the z-axis.  These lower levels are not surprising given the absence of any modeling constraints positioned along or near the axis.
In Figure \ref{fig:1390ldsbweights}, we look at the heaviest orbits making up 50\% of the total weight for a galaxy, and select those orbits which are common to both 3D luminosity density constrained and 2D surface brightness constrained models.  From the figure, it is clear that the same orbits do not have the same weights, and that just which constraints are used does influence the solution from modeling.

Schwarzschild \textit{Kin} models have low numbers of heavily weighted orbits and this requires further investigation (see the use of regularization in Section \ref{sec:reg}).  Less than 5 per cent of the orbits in these models account for 50 percent of the total weight available. Imbalances are not uncommon in M2M and Schwarzschild models but this is the most extreme of any of the models.

\subsection{Regularization}
\label{sec:reg}
As indicated in the Approach, Section \ref{sec:approach}, we investigate the use of regularization to penalize high-valued weights to see if such a constraint improves our 3D density estimates.  We only regularize Schwarzschild models as it is computationally cheap to do so: we just need to rerun the weight calculation with the regularization constraint included while M2M requires complete model reruns.  We use $2 \times 10^{-1}$ as the value of the regularization parameter \citep{Long2018}.  All the models experience an expected increase in constraint $\chi^2$ per bin values but the values remain $\leq 1$ as desired. 

\begin{figure}[h]
    \centering
    \caption{Regularization of Schwarzschild Models}
	\label{fig:schwreg}
    \begin{tabular}{cM{65mm}M{65mm}M{65mm}}
    		& \textit{3I Initial Conditions} & \textit{MDJV Initial Conditions} \\
    		\multicolumn{3}{c}{\textit{3D Density Estimation}}\\
        \textit{Nokin} & \includegraphics[width=65mm]{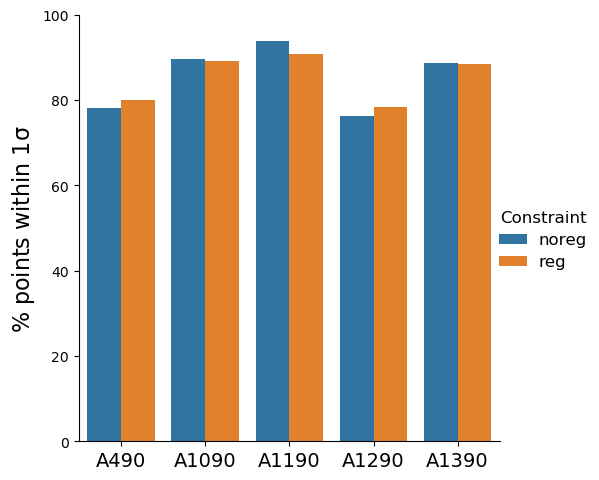}  & \includegraphics[width=65mm]{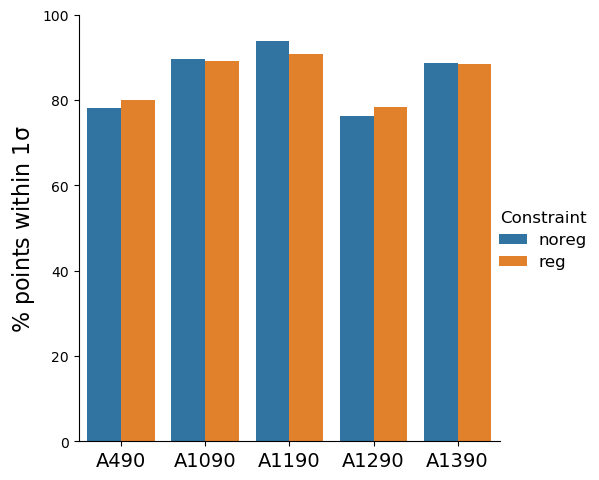} \\
        \textit{Kin}   & \includegraphics[width=65mm]{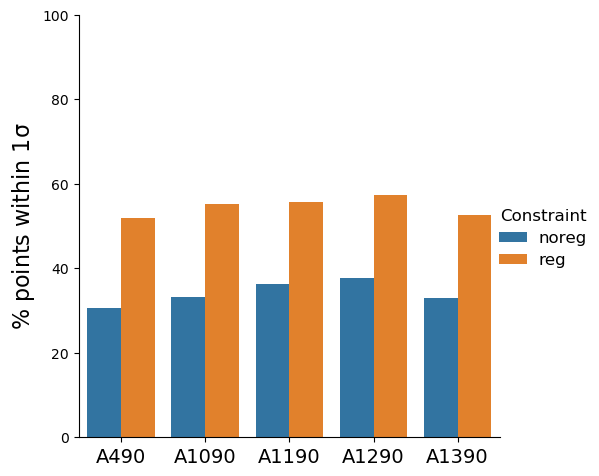}  & \includegraphics[width=65mm]{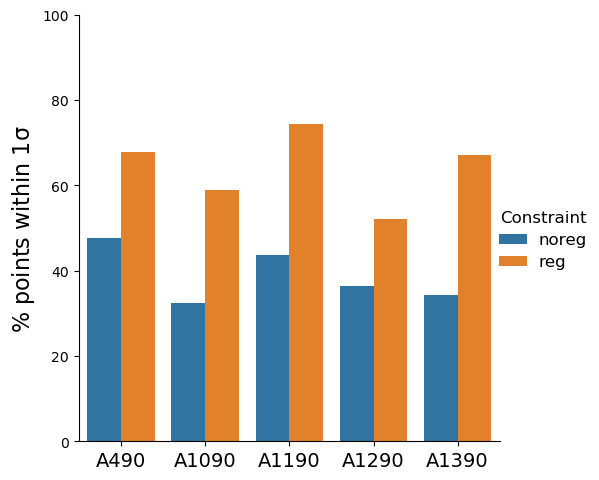} \\
        \multicolumn{3}{c}{\textit{Contributing Orbits}}\\
        \textit{Nokin} & \includegraphics[width=65mm]{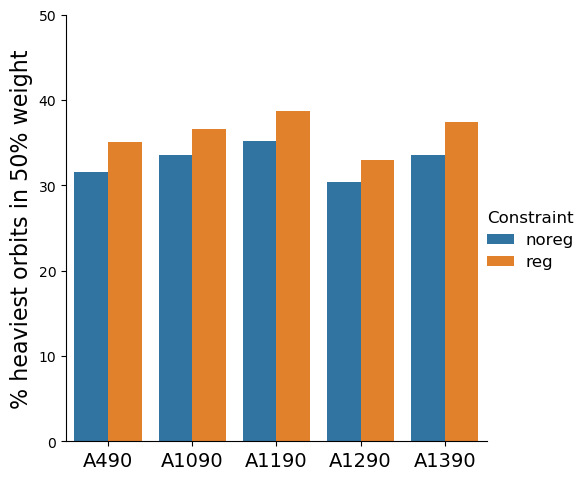}  & \includegraphics[width=65mm]{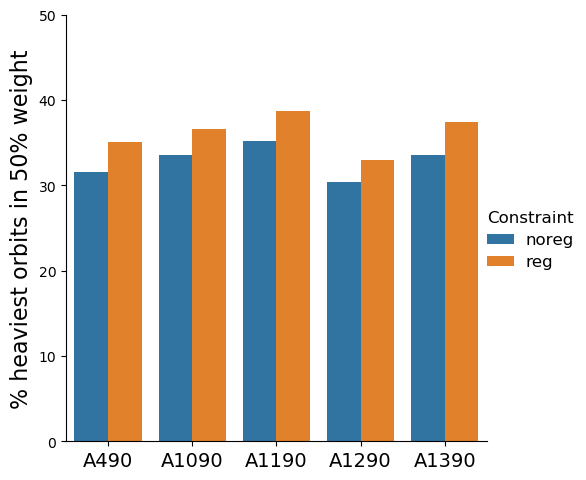} \\
        \textit{Kin}   & \includegraphics[width=65mm]{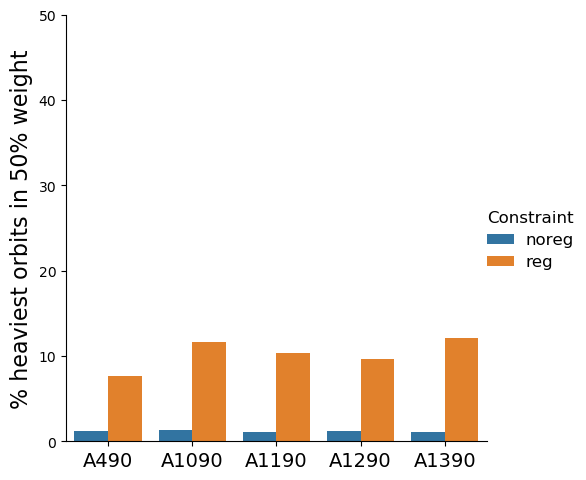}  & \includegraphics[width=65mm]{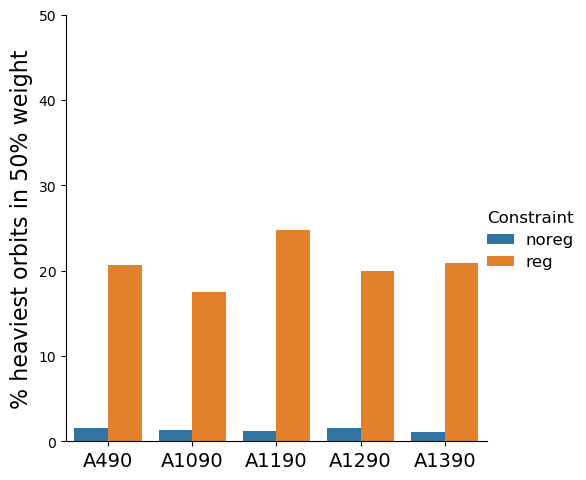} \\
    \end{tabular}

\medskip
Regularization of Schwarzschild models (Section \ref{sec:reg}) comparing results both with and without regularization.  The top two rows show the impact on 3D density estimation, and the bottom two rows the impact on the percentage of heavier orbits contributing to $50\%$ of the total weight available.  \textit{Nokin} models benefit slightly from regularization being used, the \textit{Kin} models considerably more so.
\end{figure}

We display the results in Figure \ref{fig:schwreg} for our regularized Schwarzschild \textit{Nokin} and \textit{Kin} models.  
As can be seen from Figure \ref{fig:schwreg}, regularization has a small positive benefit to the models with no kinematic constraints (the \textit{Nokin} models) but a somewhat larger effect on the models with kinematic constraints (the \textit{Kin} models).    For these models, the number of heavy orbits contributing to $50\%$ of the total weight has increased from $<2\%$ of the 8000 orbits originally used to $\approx 10\%$ for models using the 3I initial conditions, and to $\approx 20\%$ for the MDJV conditions (see the lower plots in Figure \ref{fig:schwreg}). In addition, the overall accuracy of the 3D density estimates has improved but not to a level that might be considered `good' (the upper plots in Figure \ref{fig:schwreg}).

\begin{figure}[h]
    \centering
    \caption{Regularization of Galaxy A1290 Schwarzschild \textit{Kin} Models using MDJV Initial Conditions}
	\label{fig:1290reg}
    \begin{tabular}{cc}
    		\textit{Without Regularization} & \textit{With Regularization} \\
        \includegraphics[width=70mm]{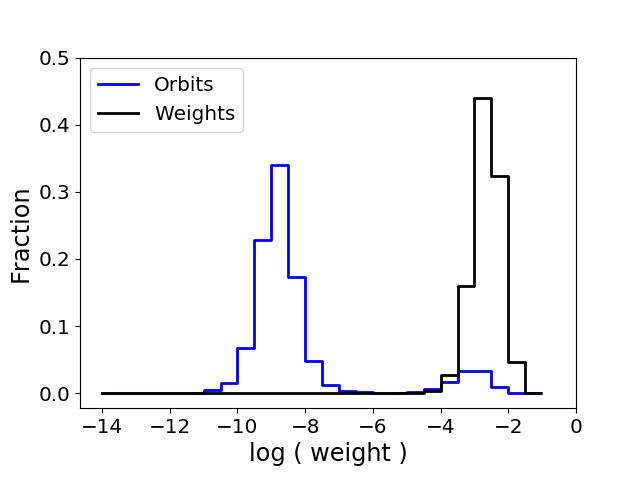}  & \includegraphics[width=70mm]{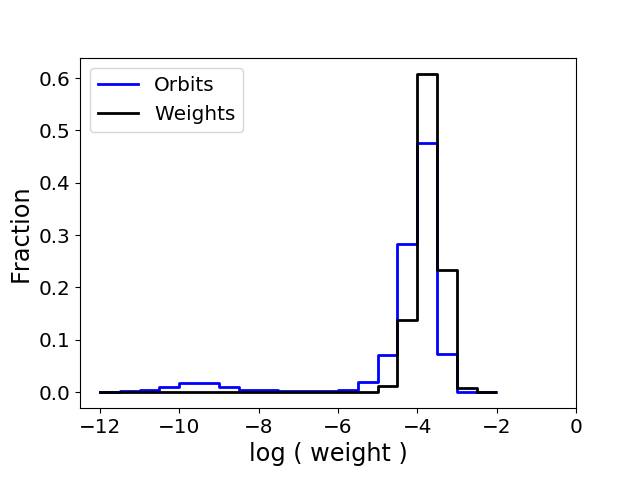} \\
        \multicolumn{2}{c}{\includegraphics[width=70mm]{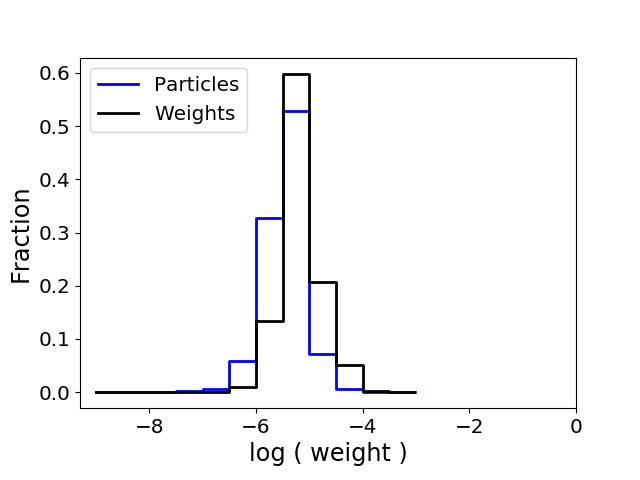}}  \\
        \includegraphics[width=70mm]{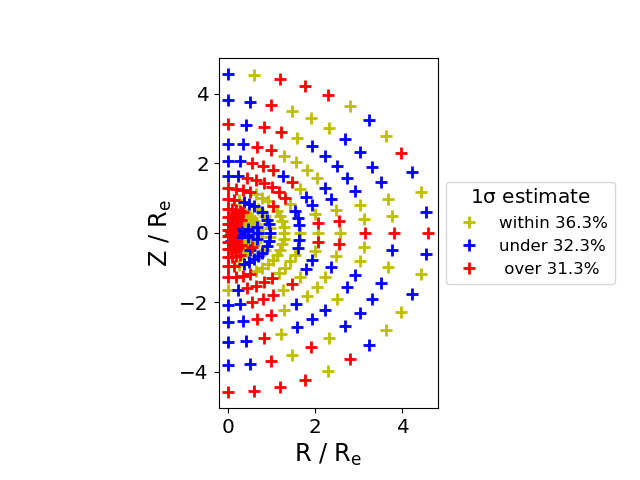}  & \includegraphics[width=70mm]{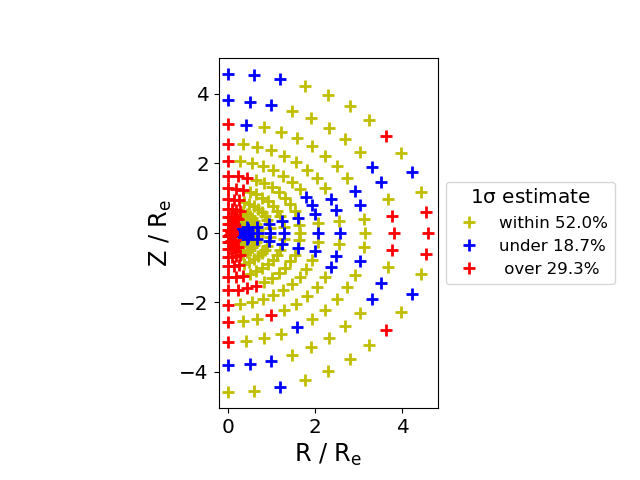} \\
    \end{tabular}

\medskip
Regularization of galaxy A1290 Schwarzschild \textit{Kin} models using MDJV initial conditions (Section \ref{sec:reg}).  The top row shows the distributions of orbits and weights. The left hand plot of the row shows the imbalance between the fractional numbers of orbits and weights when no regularization is used. The right hand panel shows how this changes when regularization is used to penalize high value weights: the imbalance is reduced as is the maximum orbit weight. The single plot in the middle row is from an unregularized M2M MDJV \textit{Kin} model of A1290 and shows a similar profile to a regularized Schwarzschild model. The bottom row shows how the how the $1\sigma$ 3D density estimate improves when regularization is used (an almost $16\%$ improvement).
\end{figure}

Regularization achieves the desired effect of penalizing high-valued weights with the maximum weight per model reducing by an order of magnitude. In Figure \ref{fig:1290reg}, we illustrate the impact this has using results from the \textit{Kin} model with MDJV initial conditions for galaxy A1290.
Using A1290 as an example, the balance between the fractional numbers of orbits and weights for individual galaxies is also improved and is similar to the unregularized M2M model.

\subsection{Inclined Models}
\label{sec:incmodels}
\begin{figure}[h]
    \centering
    \caption{Inclined \textit{Nokin} Galaxy Models - 3D Density Estimation}
	\label{fig:incmodels}
    \begin{tabular}{cM{65mm}M{65mm}M{65mm}}
    		& \textit{3I Initial Conditions}  &  \textit{MDJV Initial Conditions}  \\
        \textit{M2M} & \includegraphics[width=65mm]{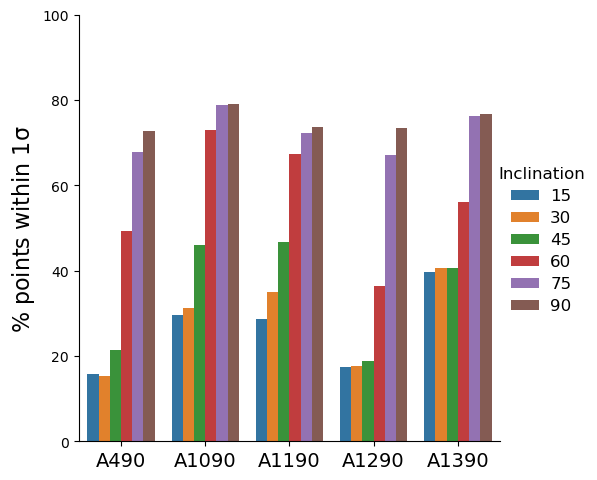}    & \includegraphics[width=65mm]{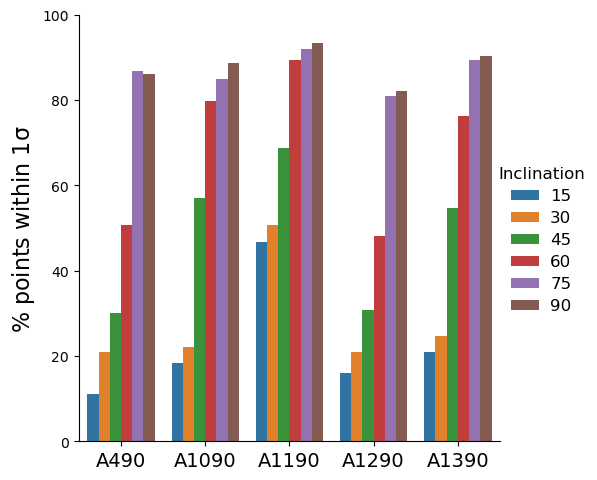} \\
        \textit{Schw} & \includegraphics[width=65mm]{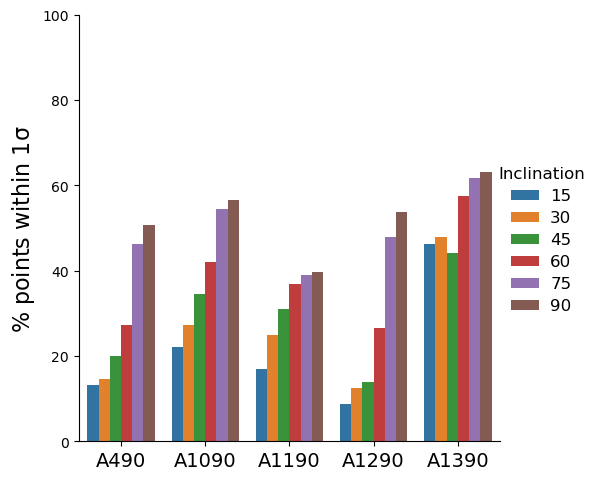}  & \includegraphics[width=65mm]{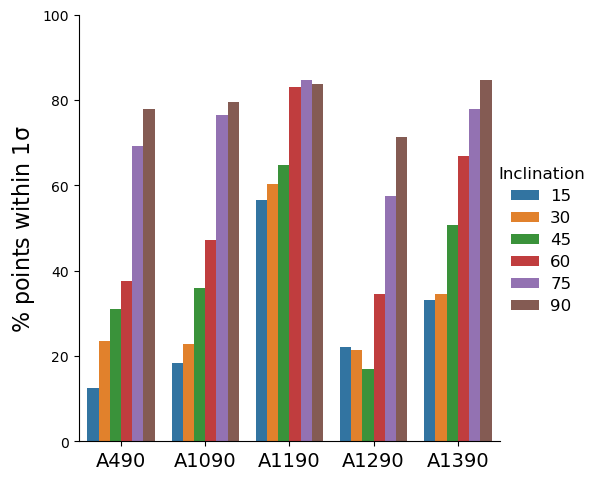} \\
    \end{tabular}

\medskip
3D density estimation using inclined \textit{Nokin} galaxy models (Section \ref{sec:incmodels}). The plots gives the percentage of points where the model 3D density agrees with the known value to within $1\sigma$ (see Eq. \ref{eqn:ptresid} with $\alpha = 1$). The M2M models using MDJV initial conditions give the best estimation percentages. Once the inclination is less than $75^\circ$, the percentages become too low for there to be much confidence in the 3D density estimates.
\end{figure}

The models and results in this section correspond to point \ref{itm:three} in the Approach, Section \ref{sec:approach}.  We choose to restrict our modeling of inclined galaxies to \textit{Nokin} models using both modeling methods and both sets of initial conditions..  Figure \ref{fig:incmodels} shows the results we obtain.  Overall, results improve as the inclination angle increases towards edge-on ($90^\circ$) with the M2M MDJV model giving the best $1\sigma$ 3D density estimation percentages. For the M2M MJDV models, the difference between the percentages for inclinations of $75^\circ$ and $90^\circ$ is typically $\approx 1\%$.  For inclinations of $60^\circ$ and $75^\circ$, the difference becomes larger and is typically $22\%$ but with a range of $5\%$ to $36\%$.  Based on these results, for inclinations lower than $75^\circ$, confidence in the 3D density estimates becomes much reduced.

As the inclination angle increases towards $90^\circ$, the percentage of the heaviest orbits contributing to $50\%$ of the total weight decreases. For M2M MDJV models the percentages are approximately $45\%$ at $15^\circ$ reducing to $35\%$ at $90^\circ$.  The Schwarzschild MDJV percentage values are about $5\%$ points lower.

\subsection{Additional Schwarzschild Models}
\label{sec:addmodels}

Based on our results so far, it is clear that Schwarzschild's method does not perform as well in our tests as the M2M method.
In this section we report on some additional Schwarzschild models we have performed for diagnostic purposes.  All the models in this section are edge-on, and cover extra orbits, longer model durations, and mixed initial conditions.

\begin{figure}[h]
    \centering
    \caption{Edge-on Models 3D Estimator Weight Imbalance}
	\label{fig:edgeonwt}
    \begin{tabular}{cM{65mm}M{65mm}M{65mm}}
    		& \textit{3I Initial Conditions}  &  \textit{MDJV Initial Conditions}  \\
        \textit{M2M}  & \includegraphics[width=70mm]{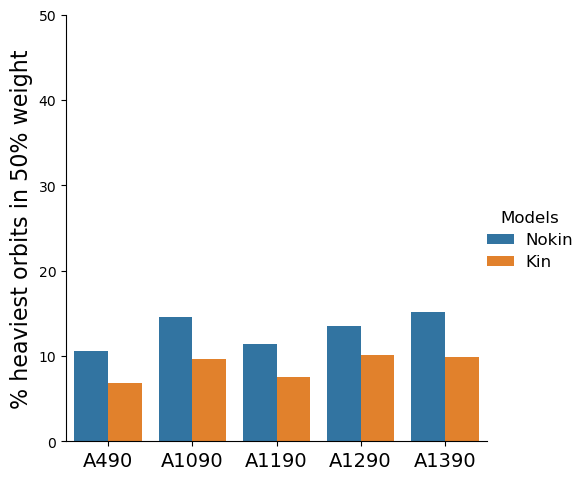}  & \includegraphics[width=70mm]{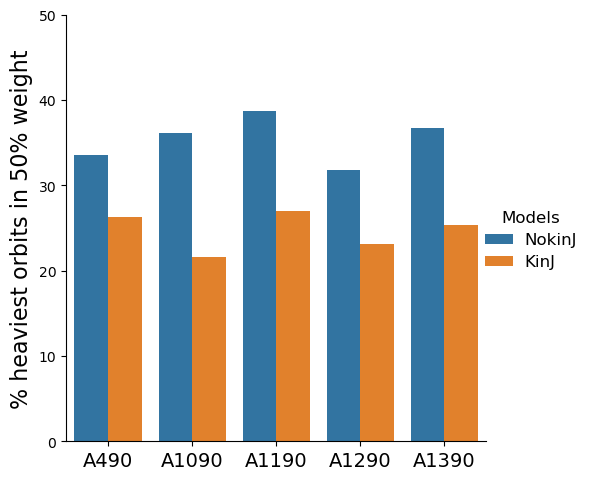} \\
        \textit{Schw} & \includegraphics[width=70mm]{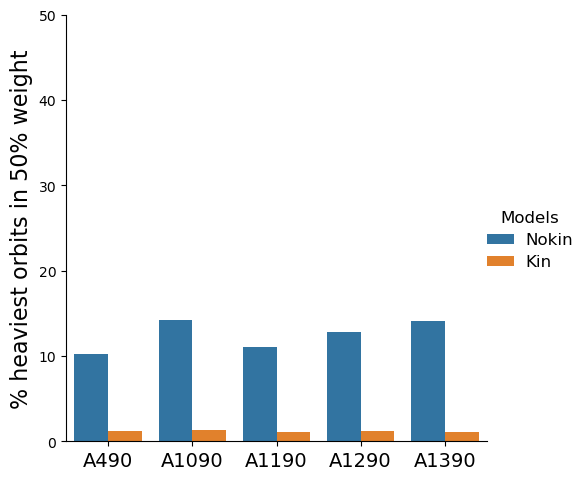} & \includegraphics[width=70mm]{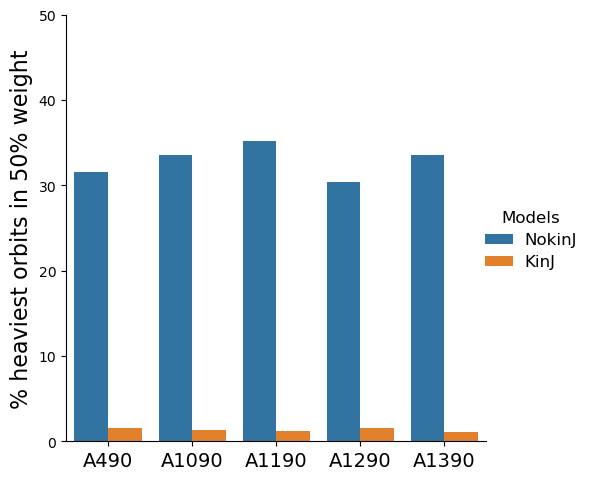} \\
    \end{tabular}

\medskip
The weight to orbit imbalance in edge-on models estimating the 3D density (Section \ref{sec:addmodels}).  The height of the bars is the percentage of orbits or particles (taking the heaviest ordered by descending weight) that account for 50 per cent of the total available. Schwarzschild \textit{Kin} models have the lowest percentages at $<2$ per cent.
\end{figure}

In Section \ref{sec:moredetail}, we identified that only a small number of orbits are really contributing to the Schwarzschild \textit{Kin} models  (see Fig. \ref{fig:edgeonwt}).  Penalizing heavy orbit weights as in Section \ref{sec:reg} does act to improve matters and causes more orbits to actively contribute.  Considering the number of heaviest particles making up 50\% of the total weight, for \textit{Nokin} models, the number is in the range $1000$ to $3000$ (out of $8000$ orbits in total). For the \textit{Kin} models the number is only $<300$, a significant drop.  It might be thought that 50\% of the weight should be supplied by 50\% of the orbits or particles but the results show this is not the case.
Increasing the number of orbits from 8000 to 16000 does not alter significantly the number of orbits contributing nor improve the 3D density estimates.  The results are displayed in Table \ref{tab:numorbit}.
\begin{table}[h]
	\centering
	\caption{Schwarzschild \textit{Kin} Models - Number of Orbits}
	\label{tab:numorbit}
		\begin{tabular}{cccccc}
		\hline
		       &           & \multicolumn{2}{c}{3I Initial Conditions} & \multicolumn{2}{c}{MDJV Initial Conditions}\\
		Galaxy &  	Total   &  Orbits       &	 3D density  &    Orbits       &	 3D density\\
		       &    Orbits  &   50\% weight  & $1\sigma$ estimate      &   50\% weight   & $1\sigma$ estimate\\
		\hline 
		A490 &	 	8000  &	   96  &  30.7\%  &  121  &  47.4\% \\
		     &      16000  &   113  &  38.2\%  &  122  &  40.3\%  \\
		A1090   &   8000  &    104  &  33.3\%  &  106  &  32.3\%  \\
		     &      16000 &    113  &  28.7\%  &  122  &  39.0\% \\
		A1190 &		8000  &	   86   & 36.3\%  &   95   &  43.7\%  \\
		     &      16000  &    92  &  19.9\%  &   123  &  50.0\% \\
		A1290    &  8000  &     98  &  37.7\%  &   122  &  36.3\% \\
		     &      16000 &    131  &  34.6\%  &   143  &   47.7\% \\
		A1390 &		8000  &	   84   &  33.0\%  &   92   &  34.3\% \\
		     &      16000  &    88   &  33.1\%  &   102  &  39.0\% \\
		\hline
	\end{tabular}

\medskip
By default, we use 8000 orbits in all Schwarzschild models.  Increasing the number to 16000 has little impact on the number of orbits actively contributing to matching the observable constraints and only marginally improves the 3D density estimates (Section \ref{sec:addmodels}). The number of the heaviest orbits making up 50\% of the total weight is shown in the columns labeled `Orbits 50\% weight'.  The 3D density columns give the percentage of 3D density point estimates with $1\sigma$ of the known values.
\end{table}
It is clear that the number of contributing orbits does not scale with the total number of orbits used.  It would therefore seem that, once a sufficiently varied pool of orbits is available, the number contributing is probably related to the competing demands of the constraining kinematic and luminosity observables.  This result is consistent with \citet{Jin2019} where a different Schwarzschild implementation was used. The findings from Section \ref{sec:reg} do appear to offer an alternative strategy which is to use MDJV initial conditions with regularization penalizing high value weights.

Schwarzschild models typically have shorter modeling durations than M2M models: in M2M, time has to be allowed for particle weight convergence.  In this section, we report on the effect of doubling the Schwarzschild durations from 50 to 100 and 200 hmdtu.  Our results for the \textit{Nokin} set of models are shown in Table \ref{tab:extruntime}.
\begin{table}[h]
	\centering
	\caption{Schwarzschild \textit{Nokin} Models - Extended Run Times}
	\label{tab:extruntime}
		\begin{tabular}{cccc}
		\hline
		       &          & 3I Initial Conditions & MDJV Initial Conditions\\
		Galaxy & Duration & 	  3D density  &   3D density  \\
		       & (hmdu)     & $1\sigma$ estimates   &  $1\sigma$ estimates \\
		\hline 
		A490 & 50 & 50.0\% & 73.5\%  \\
 			& 100 & 59.6\%  & 78.7\%  \\
 			& 200 & 64.0\%  & 79.4\%  \\
		A1090 & 50 & 56.6\%  & 81.6\%  \\
 			& 100 & 57.4\%  & 83.1\%  \\
 			& 200 & 62.5\%  & 80.1\%  \\
 		A1190 & 50 & 39.7\%  & 82.4\%  \\
			& 100 & 41.2\%  & 82.4\%  \\
 			& 200 & 43.4\%  & 83.8\%  \\
 		A1290 & 50 & 53.7\%  & 69.1\%  \\
			& 100 & 60.3\%  & 72.1\%  \\
 			& 200 & 66.2\%  & 72.1\%  \\
 		A1390 & 50 & 63.2\%  & 81.6\%  \\
			& 100 & 66.9\%  & 83.1\%  \\
 			& 200 & 66.9\%  & 83.8\%  \\
		\hline
	\end{tabular}

\medskip
By default, we use durations of 50 hmdtu for all Schwarzschild models.  Increasing the duration to 100 and 200 hmdtu may yield small improvements in the 3D density estimates (Section \ref{sec:addmodels}). The two 3D density columns give the percentage of 3D density point estimates within $1\sigma$ of the known values ($\alpha$ is 1 in Eq. \ref{eqn:ptresid}).
\end{table}
As can be seen, the point estimate percentages may show small improvements, particularly with the 3I initial conditions but less so with the MDJV conditions.

We have tried mixed initial conditions where we take some combination of 3I and MDJV initial conditions.  As might be expected, we find that the results lie between the two extremes represented by the 3I and MDJV schemes.  The closer the mixed conditions are to a full MDJV set the better the 3D luminosity density estimates become.

\subsection{Computer Utilization}
\label{sec:computil}
The software implementation used is based on the designs in \cite{Long2010, Long2012} and as re-implemented in Python 3 and Cython for \cite{Long2016} and \cite{Long2018}.  Both Schwarzschild's method and the M2M method are supported by the same implementation.  The main modeling software is written in Cython with some functions in C for performance reasons. The implementation has been parallelized with the Message Passing Interface (MPI) being used during model execution, and with OpenMP during preparation and analysis.  
Modeling runs were executed on an HPC cluster for M2M, and a desktop PC for Schwarzschild's method.  As an approximate guide to elapsed times, based on modeling for galaxy A1290, M2M runs on 28 processing cores took 65min for models not using kinematic constraints, and 100min for models using kinematics.  Schwarzschild runs with no kinematics took 11min on 8 cores to create the orbit library with a further 2min to determine the orbit weights, while runs with kinematics took 18min for the library and 4min for the weights. Increasing the number of orbits with Schwarzschild's method as in Section \ref{sec:addmodels} does not result in a simple linear increase in elapsed time in determining the weights: doubling the orbits from 8000 to 16000 meant 25min were required for the weights.

\section{Discussion}
\label{sec:discussion}
In this section, we review qualitatively our results from Section \ref{sec:results}, and comment on their implications for modeling and for understanding the 3D structures and distributions in external galaxies.  In particular, as described in the Introduction, we wish to understand whether the M2M and Schwarzschild methods are able to accurately deproject an edge-on axisymmetric galaxy's 2D luminosity density to give its exact underlying 3D density.

Both M2M and Schwarzschild galaxy models meet supplied 3D luminosity density constraints, no matter which initial conditions scheme is used or whether the modeling is with or without kinematic constraints (Section \ref{sec:3Dmodels}).
This is an important first step: if it had not been met, the value of continuing with the research would have been doubtful.  In these initial tests, Schwarzschild's method performed marginally better in meeting the 3D constraints than M2M, and the more usual three integral scheme (3I) for initial conditions was outperformed by the observationally based scheme (MDJV). 

Using 2D luminosity constraints, the performance of the MDJV scheme is re-affirmed, but the roles of M2M and Schwarzschild modeling are reversed with M2M leading in its ability to estimate the 3D density (Section \ref{sec:2Dmodels}). Modeling close to the z-axis, or in the center of the galaxy is a problem particularly for Schwarzschild's method.  Also, the accuracy of the 3D estimates drops for both methods with the introduction of kinematic constraints, markedly so for Schwarzschild's method.

Analyzing and comparing the orbit (particle) weights between models using 3D and models using 2D luminosity constraints, it is clear that the weight determination mechanisms prefer orbits that can contribute towards meeting the constraints (Section \ref{sec:moredetail}).  It is not surprising therefore that Schwarzschild's method is unable to produce a `good' estimate of the 3D density when it is not constrained to do so.  Low $|\lambda_z|$ orbits are crucial to reproducing observations or producing estimates along or close to the z-axis and the center of the galaxy.  This gives some insight into why the MDJV initial conditions are more effective than the 3I initial conditions: the MDJV orbit mix is better suited to the models we have run.
It must be stressed that orbit weightings do not indicate the importance of orbits in an astrophysical sense only their usefulness to a numerical optimization algorithm. Different initial conditions mean different orbits are used resulting in different solutions.  This means that any orbital analysis work may not be truly representative of a galaxy as is all work using phase space coordinates arising in the modeling processes.  The constraints used also influence the relative importance of the orbits in the weighting process with, for example, low $|\lambda_z|$ orbits receiving higher weights when a 3D density constraint is employed rather than 2D surface brightness.

The low active orbit numbers for Schwarzschild models with kinematic constraints can be alleviated by using regularization to penalize heavy orbit weights (Section \ref{sec:reg}).  The 3D density estimates improve as a result of regularization being using  but not to the extent that the target values are well-reproduced.  Other approaches, such as more orbits, or longer model durations, or mixed initial conditions, to improve the estimates have minimal impact (Section \ref{sec:addmodels}).  Overall, it was not been possible to cross the gap to the results achieved by the M2M method.  Perhaps more consideration of other aspects or features, for example, whether or not any mass to velocity dispersion anisotropy \citep{Gerhard1998} is influencing the results,  is required.

The main thrust of our work is to investigate the accuracy of density estimation for edge-on galaxies.  Some modeling runs with inclined galaxies were performed, and we found that M2M MDJV models appear to produce approximately the same 3D density estimates for inclinations between $75^{\circ}$ and $90^{\circ}$ (edge-on) (Section \ref{sec:incmodels}).  Below $75^{\circ}$ the accuracy of density estimates drops significantly.

From our results and the summary above, it appears that both Schwarzschild's method and the M2M method are not able to recreate accurately 3D density distributions of edge-on axisymmetric galaxies from 2D data. The processes within the two methods do not support \cite{Rybicki1987}: they are designed as replicators of galaxy data not as 2D to 3D deprojectors.
More generally, these modeling methods can not be used to understand accurately any 3D structures and distributions inside an external galaxy from 2D data: they are able to indicate, to illustrate,  what a galaxy might be like internally but their models cannot be taken as accurate or definitive.  This is no more than can be expected from the external galaxy data astronomers are currently capable of collecting. The methods are able to meet an imposed axisymmetric 3D luminosity density constraint and this may be an acceptable assumption to make, provided of course that it is clearly stated.  Appropriate assumptions for other physical, less smooth, 3D distributions (for example, age or metallicity) are much less clear, and, in addition, it must be remembered that such distributions are modeled as luminosity weighted quantities and the accuracy of their modeling depends on an accurate 3D luminosity density being produced first.

We have considered whether `usable 3D regions' might exist, for example close to the equatorial plane, or avoiding the z-axis.  The M2M \textit{Kin} models using MDJV may offer such a solution provided the unusable regions can be consistently identified without knowing what the true 3D distribution is.
Imposing some so called `strong priors' on the modeling may be appropriate but can not fundamentally change any issues: the projection / deprojection issues arising from the observed data are not surmountable.  Extrapolating, it is not obvious how 3D distributions from galaxies in cosmological simulations can be validated. 

From a broader perspective, accurate estimation of a galaxy's 3D luminosity density distribution has an important role to play in estimating other dynamical quantities such as the central dark matter fraction or distribution in a galaxy, or a galaxy's mix of high to low angular momentum orbits.  Perhaps, solutions to these matters lie elsewhere, through exploitation of machine learning techniques, for example.

Recently, there have been papers published which would have benefited from the inclusion of caveats (as indicated above) on their results.  Rather than dwell on these negative aspects, we encourage researchers to follow the lead in \cite{Jin2019, Jin2020} and make the limitations of their work clearly visible to avoid misleading the less experienced in our astronomical community.

Our Schwarzschild modeling activities benefited operationally from the use of a multi-core workstation, avoiding the need to utilize HPC systems (Section \ref{sec:computil}).  
Development of an efficient GPU workstation M2M implementation would enable the same benefits to be enjoyed by M2M modeling activities.

Only one software implementation has been used in this research, and it is likely that different implementations will achieve slightly different results. We believe that our analysis is sound and is a reflection of the capabilities of the methods, and is not implementation specific.   We recommend however that all implementation owners should check the 3D behaviors of their implementations as part of presenting or publishing any 3D results.

\section{Conclusions}
\label{sec:conclusions}
We have met the objectives set out in the Introduction, Section \ref{sec:intro}.  Based on our results in Section \ref{sec:results} and discussed in Section \ref{sec:discussion}, we now understand that Schwarzschild's method and the M2M method are unable to recover accurately 3D luminosity distributions from their 2D projections for our sample of axisymmetric edge-on galaxies constructed from a cosmological simulation.  The methods do not support the theory in \cite{Rybicki1987}, and no assumptions to this end should be made.  

Our results reinforce the messages in \cite{Jin2019, Jin2020} that Schwarzschild models of external galaxies (and M2M models as well) can only be thought of as illustrative and not as definitive, accurate statements.  The projection / deprojection issues arising from observations of external galaxies are not surmountable using these stellar dynamical modeling methods. With adequate training data, it may be that more insight into the 3D structures of external galaxies might be forthcoming using machine learning techniques. Our results have implications broader than just luminosity density, and affect other luminosity-weighted distributions within galaxies, for example, age and metallicity.

\begin{acknowledgements}
The authors thank Shude Mao for his contribution at the inception of the project, and acknowledge the role of Yan Liang in the early stages of galaxy data preparation from the IllustrisTNG 100 simulation.  This work is partly supported by the National Key Basic Research and Development Program of China (No. 2018YFA0404501 to Shude Mao), and by the National Science Foundation of China (Grant No. 11821303, 11761131004 and 11761141012 to Shude Mao).  M2M modeling runs were performed on the \textit{Orion} high performance computer cluster of the Department of Astronomy at Tsinghua University.
\end{acknowledgements}

\appendix

\section{Plots of Galaxy Data}
\label{app:gdata}
\begin{figure}[h]
    \centering
    \caption{Galaxy Data}
	\label{fig:stellarmge}
    \begin{tabular}{cM{45mm}M{45mm}M{45mm}M{45mm}}
    		 & Log Surface Brightness & Surface Brightness MGE & Mean LOS Velocity \\
        A490  & \includegraphics[width=45mm]{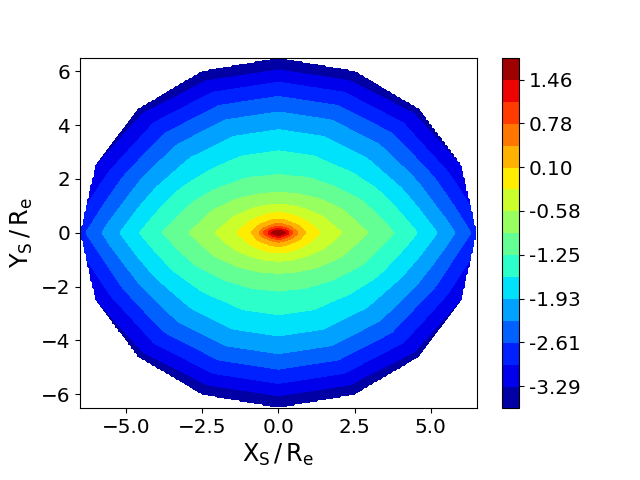} & \includegraphics[width=45mm]{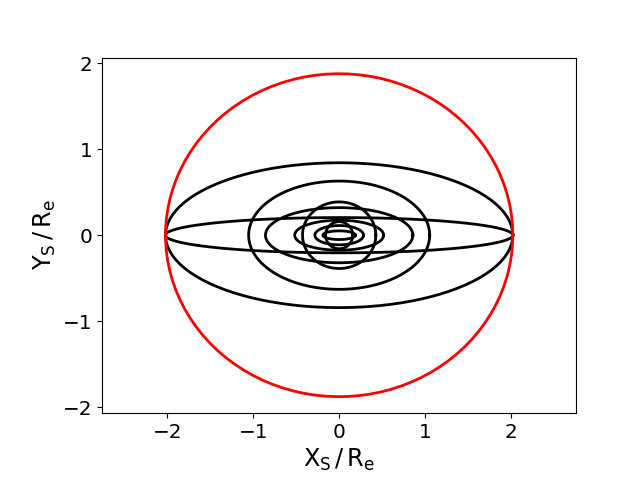} & \includegraphics[width=45mm]{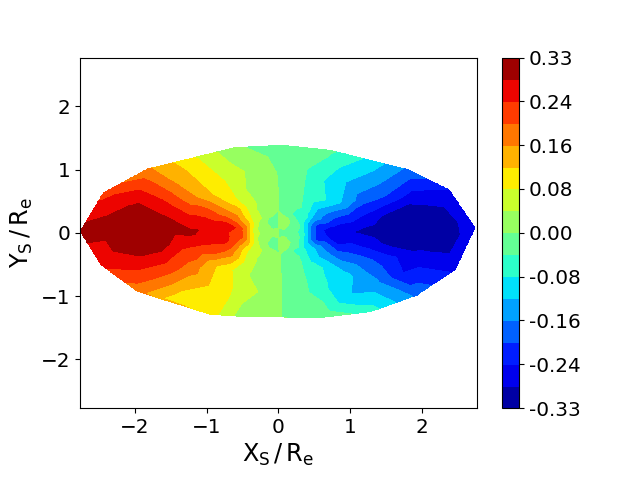} \\
        A1090 & \includegraphics[width=45mm]{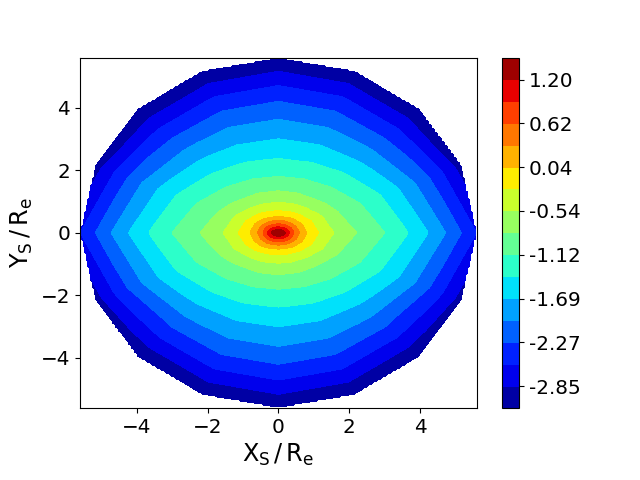} & \includegraphics[width=45mm]{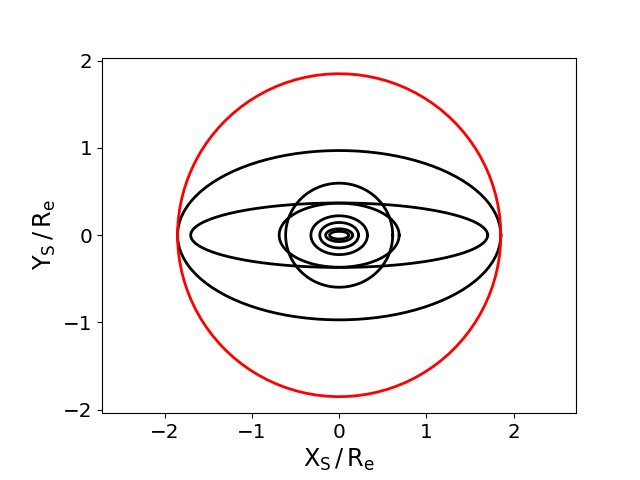} & \includegraphics[width=45mm]{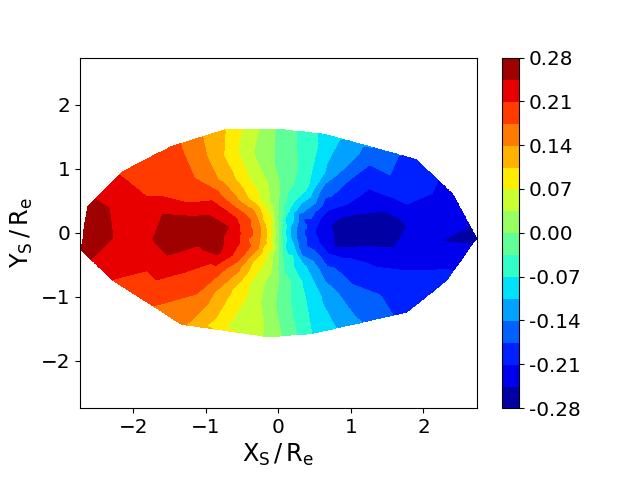} \\
        A1190 & \includegraphics[width=45mm]{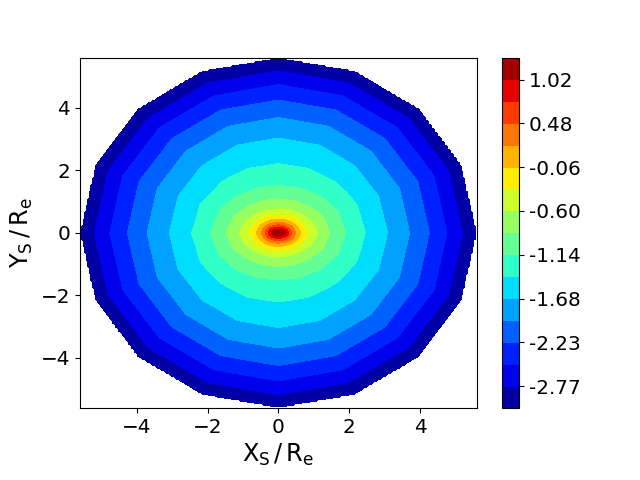} & \includegraphics[width=45mm]{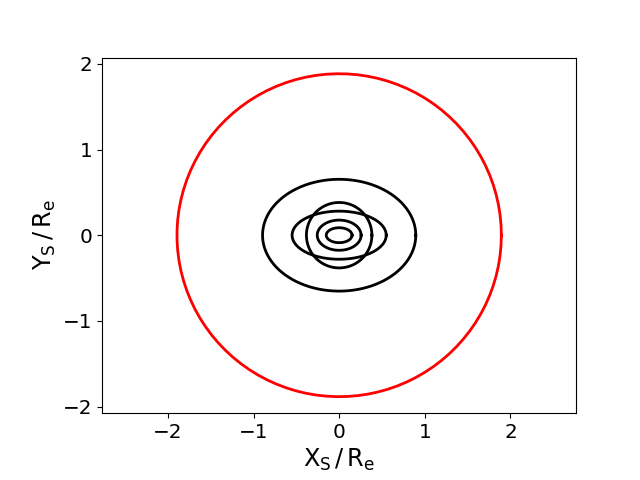} & \includegraphics[width=45mm]{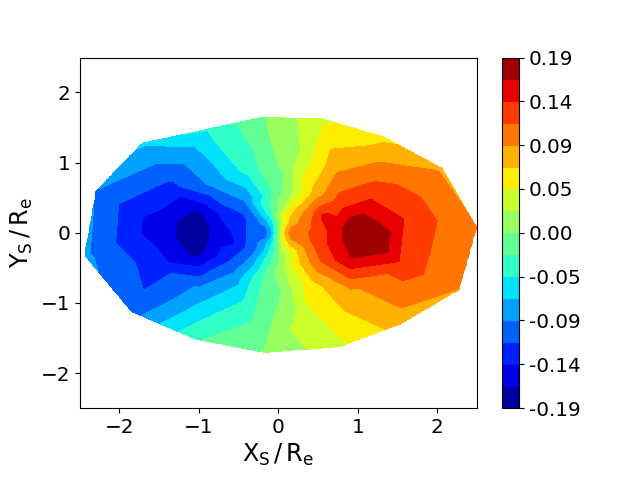} \\
        A1290 & \includegraphics[width=45mm]{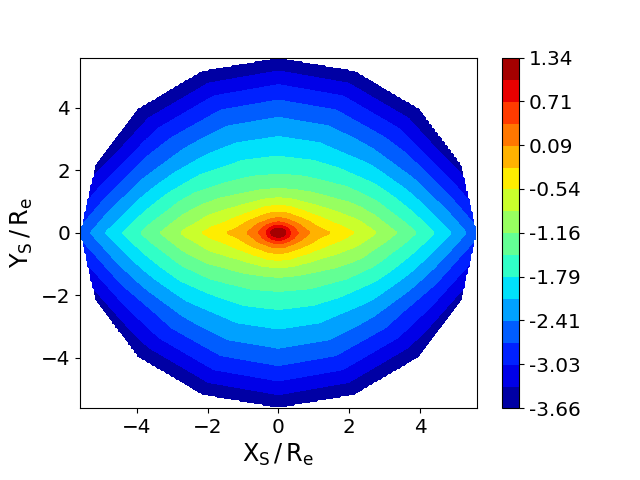} & \includegraphics[width=45mm]{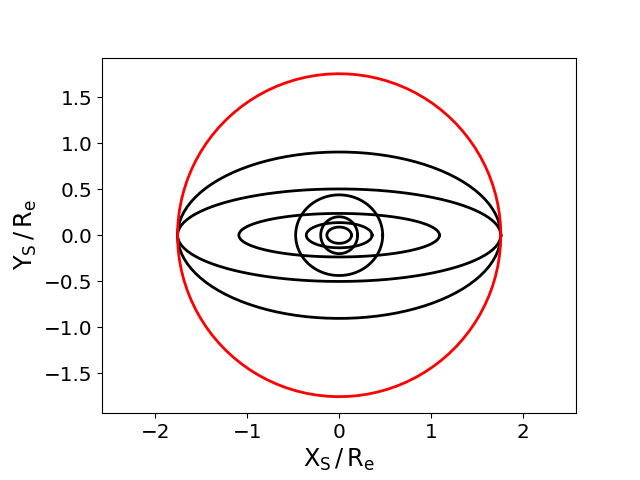} & \includegraphics[width=45mm]{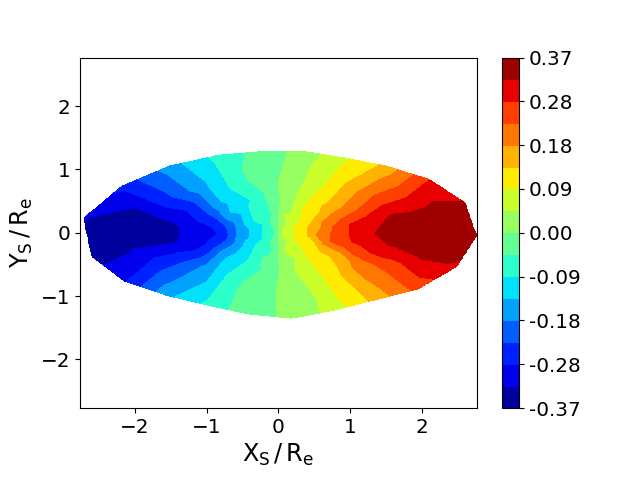} \\
        A1390 & \includegraphics[width=45mm]{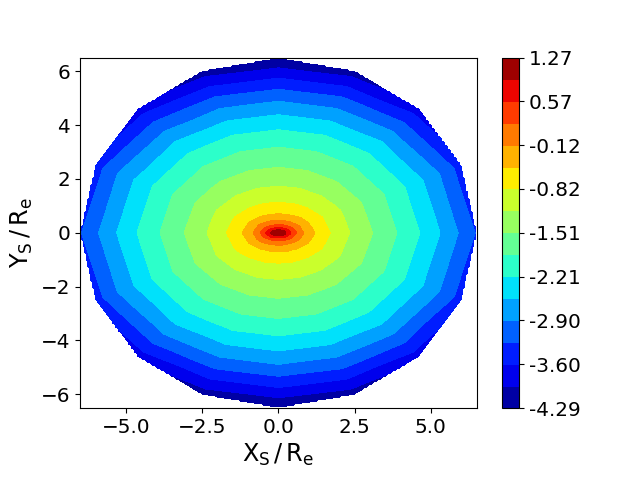} & \includegraphics[width=45mm]{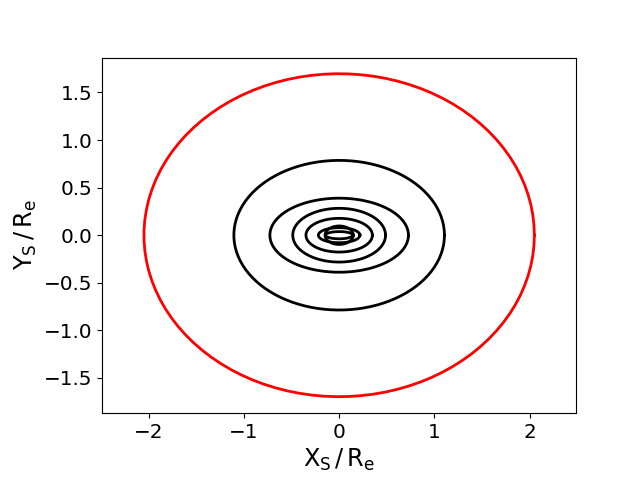} & \includegraphics[width=45mm]{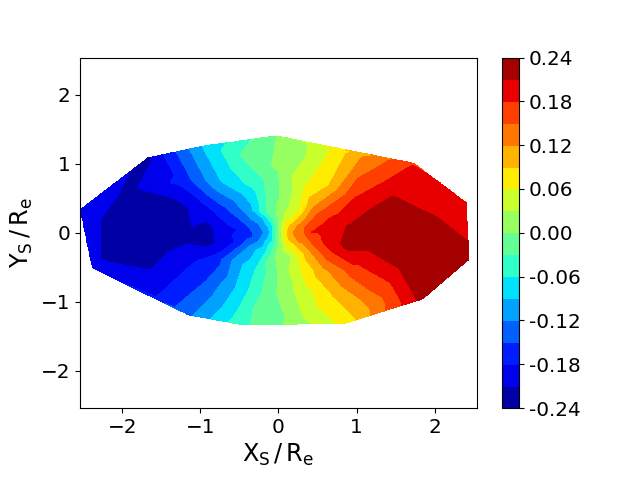} \\
    \end{tabular}
    
\medskip  
Galaxy plots for the sample galaxies.  The columns are ($\log$) surface brightness, the surface brightness MGE Gaussians, and the mean line-of-sight (los) velocity.   Each Gaussian in an MGE is plotted with a major axis length of one Gaussian sigma.  The red outer Gaussian is used for determining the galaxy sizes for modeling.  The internal units used in the plots are as described in Section \ref{sec:common}.
\end{figure}

In this Appendix, in Figure \ref{fig:stellarmge}, we show plots of some of the data from the galaxies we model.   

\bibliographystyle{raa}
\bibliography{ms2020-0474}

\end{document}